\pdfoutput=1
\documentclass[a4paper, twoside, 12pt]{article}
\usepackage[utf8]{inputenc}
\usepackage[T1]{fontenc}
\usepackage{longtable}
\usepackage{hyperref}
\usepackage{caption}
\usepackage{amsmath}
\usepackage{bm}
\usepackage{amssymb}
\usepackage{wrapfig}
\usepackage{algorithm}
\usepackage{pbox}
\usepackage{algcompatible}
\usepackage{titlesec}
\usepackage{array}
\usepackage{graphicx}
\usepackage{subcaption}
\newcommand\simpleparagraph[1]{%
  \stepcounter{paragraph}\paragraph*{\theparagraph\quad{}#1}}
\titleformat{\paragraph}
{\normalfont\normalsize\bfseries}{\theparagraph}{1em}{}
\titlespacing*{\paragraph}
{0pt}{3.25ex plus 1ex minus .2ex}{1.5ex plus .2ex}

\DeclareMathOperator*{\argmax}{arg\,max}
\DeclareMathOperator*{\argmin}{arg\,min}
\usepackage[left=1.5cm, right=3cm, top=3cm, bottom=3cm, bindingoffset=1.5cm, head=15pt]{geometry} 
\usepackage{setspace}
\usepackage{fancyhdr}
\usepackage[numbers]{natbib}
\title{Machine Learning for Genomic Data}

\newcommand{\thesisauthor}{Akankshita Dash} %input name
\newcommand{\thesistype}{Undergraduate Thesis } % Set either to Bachelor or Master
\newcommand{\supervisor}{Assoc. Prof. Vincent Tan}
\newcommand{\cosupervisor}{Dr. Huili Guo}

\begin{document}

%%%%%%%%%%%%%%%%%%%%%%%%%%%%%%%%%%%%%%%%%%%%%%%%%%%%%%%%%%%%%
%TITLE PAGE (Pre-defined, just change parameters above)
%%%%%%%%%%%%%%%%%%%%%%%%%%%%%%%%%%%%%%%%%%%%%%%%%%%%%%%%%%%%%
%%%%%%%%%%%%%%%%%%%%%%%%%%%%%%%%%%%%%%%%%%%%%%%%%%%%%%%%%%%%%
%TITLE PAGE
%%%%%%%%%%%%%%%%%%%%%%%%%%%%%%%%%%%%%%%%%%%%%%%%%%%%%%%%%%%%%
\makeatletter
\begin{titlepage}
    \begin{center}
        \vspace*{1cm}

        \begin{figure}[htbp]
             \centering
             \includegraphics[bb=0 0 242 118]{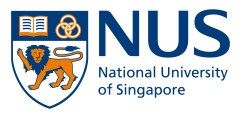}
        \end{figure}
        
        \Large
        \textbf{\@title}

        \vspace{1.5cm}
        
        \thesistype{}
        
        \vspace{1cm}

        \vspace{1cm}

        \large
        \textbf{Author}: \thesisauthor{} \\
        \large
        \textbf{Supervisor}: \supervisor{}\\
        \large
        \textbf{Co-Supervisor}: \cosupervisor{}
        \vspace{1.5cm}

\textit{        A thesis submitted in partial fulfilment of the requirement for the degree of
Bachelor of Science with Honours in Applied Mathematics}
        \vspace{1.5cm}
        
        \large
        Department of Mathematics\\
        Faculty of Science\\
        National University of Singapore\\

        \vspace{1cm}
        \textbf{8th March, 2019}

    \end{center}
\end{titlepage}
\makeatother

%%%%%%%%%%%%%%%%%%%%%%%%%%%%%%%%%%%%%%%%%%%%%%%%%%%%%%%%%%%%%
%SOOA
%%%%%%%%%%%%%%%%%%%%%%%%%%%%%%%%%%%%%%%%%%%%%%%%%%%%%%%%%%%%%
\clearpage
\thispagestyle{empty}
\section*{Acknowledgements}
\label{sec:Acknowledgements}

First and foremost, I would like to express my utmost gratitude to my supervisors; \textbf{Assoc. Professor Vincent Tan}, for introducing me to machine learning in MA4270, and \textbf{Dr. Huili Guo} for breaking down complex bioinformatics concepts into simple ones for me to understand. I am extremely grateful for their guidance during the entire period of my honours project. Throughout the year, they have been understanding of my strengths and weaknesses and tolerant of my mistakes, guiding me along when I hit roadblocks, all the while suggesting new ideas and patiently helping me clarify my doubts. I have learnt so much from both of them, both about academia and life, and without their constant guidance, this thesis would not have been possible. I couldn't have asked for better mentors. Thank you.

I would also like to thank the people in Dr. Huili Guo's lab, especially Indrik and Shawn, for being so helpful when I was struggling with certain concepts. I also express my appreciation to Zhao Yue for being a helpful companion on this journey, and wish him well for his FYP.

To my friends - Spatika, Srishti, Megha, Bushra and Kai Ting - thank you for your support and encouragement. You have listened to my never-ending rants, put up with my idiosyncrasies, seen me at my lowest points and yet still stuck by me.

Last but not the least, I would like to thank my family. Despite living so far, you have always managed to be there for me. I'm so grateful for your support and encouragement throughout my undergraduate years. To my sister and brother-in-law; despite your busy lives and the personal tragedy you faced last year, you continued lending a listening ear to my insignificant complaints. To my Mom and Dad - you have always been my pillars of support.You have never pressured me and always encouraged me to do my best. From sending me home-cooked meals, visiting me in Singapore when I couldn't make it home, and offering me life advice when I was at crossroads, you have helped me up when I felt like I was on the edge of failure. Thank you so much.

\vspace{1.5cm}
\noindent
\textbf{\thesisauthor{}}

%%%%%%%%%%%%%%%%%%%%%%%%%%%%%%%%%%%%%%%%%%%%%%%%%%%%%%%%%%%%%
%ABSTRACT
%%%%%%%%%%%%%%%%%%%%%%%%%%%%%%%%%%%%%%%%%%%%%%%%%%%%%%%%%%%%%
\clearpage
\thispagestyle{empty}
\section*{Abstract}

This report explores the application of machine learning techniques on short time-series gene expression data. Although standard machine learning algorithms work well on longer time-series', they often fail to find meaningful insights from fewer timepoints. 

In this report, we explore model-based clustering techniques. We combine popular unsupervised learning techniques like K-Means, Gaussian Mixture Models, Bayesian Networks, Hidden Markov Models with the well-known Expectation Maximization algorithm. K-Means and Gaussian Mixture Models are fairly standard, while Hidden Markov Model and Bayesian Networks clustering are more novel ideas that suit time-series gene expression data.

%%%%%%%%%%%%%%%%%%%%%%%%%%%%%%%%%%%%%%%%%%%%%%%%%%%%%%%%%%%%%
%TOC,TOF,TOT
%%%%%%%%%%%%%%%%%%%%%%%%%%%%%%%%%%%%%%%%%%%%%%%%%%%%%%%%%%%%%
\clearpage
\pagenumbering{Roman}
\tableofcontents

\pagenumbering{arabic}

%%%%%%%%%%%%%%%%%%%%%%%%%%%%%%%%%%%%%%%%%%%%%%%%%%%%%%%%%%%%%
%MAIN PART
%%%%%%%%%%%%%%%%%%%%%%%%%%%%%%%%%%%%%%%%%%%%%%%%%%%%%%%%%%%%%

% SEC1
\clearpage
\section{Introduction}
\label{sec:Intro}

\textbf{Machine learning} is a field of artificial intelligence that uses statistical techniques to give a computer the ability to learn from data, without explicitly programming it to do so. 

\textbf{Genomics} is a branch of molecular biology concerned with the structure, function, evolution, and mapping of an organism's complete set of DNA.

In this project, we apply machine learning techniques to \textit{viral} genomic data recovered from human cells, and attempt to derive useful insights about the \textbf{EV-A71 virus}, also known as hand-foot-and-mouth disease (HFMD).

\subsection{Machine Learning}

With the large volume of data available nowadays, it is impossible to apply statistical techniques and derive useful insights without making use of a computer. 

Considered one of the biggest innovations since the microchip, machine learning enables humans to build models and capture patterns within data. It allows one to define simple rules to improve the performance of these models. The applications of machine learning are manifold - from spam filters in e-mail to online banking, it is omnipresent. Machine learning's presence is growing, and its use cases continue increasing.

A machine learning model can be defined by its training data and the set of hypotheses functions.

Machine learning is broadly divided into the following 3 categories:-
\begin{itemize}
    \item Supervised Learning
    \item Unsupervised Learning
    \item Reinforcement Learning
\end{itemize}

In \textbf{supervised learning}, the training data are labelled, i.e. each data point is attached with its true value, or \textit{ground truth}. Typical problems include \textit{classification} (binary/multi-class) and \textit{regression}. E.g. If given a training dataset consisting of pictures of cats and dogs, our model accurately predicts whether a new image is that of a cat or a dog. This is a \textit{classification} problem. Another simple example is of a dataset with people's heights and weights. Given a new person's height, we should be able to accurately predict weight. This is a \textit{regression} problem.

In \textbf{unsupervised learning}, unlike supervised learning, the training data are unlabelled. The number of labels may or may not be specified beforehand. Unsupervised learning is commonly used as an initial exploratory technique to draw useful patterns from the dataset. E.g. Given many different books, classifying them into different categories, or \textit{clusters}, can help us identify the genres they belong to. 

\textbf{Reinforcement learning} is different from both supervised and unsupervised learning. It focuses on maximising reward potential from the data by balancing the trade-off between exploration and exploitation. E.g. In recommendation engines, the items showed to a particular user may be ordered based on how much they liked previous products or how \textit{different} they are from previous products. This is done to maximise the number of clicks, i.e. the \textit{reward}.

Since the data that we have in this project are unlabelled, our focus is on \textit{unsupervised learning}.

\subsection{Genomics}

Genomes contain the complete set of genes present in an organism, and genomics describe how this genetic information is stored and interpreted in the cell.

\subsubsection{Key Definitions and Terms}
\begin{itemize}
    \item \textbf{Nucleobases}: Components of nucleotides, that are the building blocks of DNA - Adenine (A), Thymine (T), Cytosine (C), Guanine (G). Additionally, Uracil (U) is used instead of Thymine (T) for RNA.
    \item \textbf{Codons}: A sequence of three DNA or RNA nucleotides that corresponds with a specific amino acid or stop signal during protein synthesis. For eukaryotes, there are a total of 64 such codons, which map to 20 amino acids (and stop signals) through the \textit{genetic code}.
    \item \textbf{Amino acid}: Organic molecules that make up the building block of proteins. There are 20 canonical amino acids.
    \item \textbf{DNA}: Deoxyribonucleic acid, a double helix molecule, that contains the genetic material for all organisms on Earth (including viruses).
    \item \textbf{RNA}: Ribonucleic acid, a single helix, polymeric molecule that is essential in various biological roles in coding, decoding, regulation and expression of genes. Biologically active RNAs, including transport, ribosomal and small nuclear RNA (tRNA, rRNA, snRNAs) fold into unique structures guided by complementary pairing between nucleotide bases.
    Cellular organisms use messenger RNA (mRNA) nucleotides to direct synthesis of specific proteins.
    \item \textbf{Protein}: Sequences of amino acids that carry out the majority of cellular functions such as motility, DNA regulation, and replication.
    \item \textbf{Transcription}: The process in which a DNA sequence of a gene is rewritten, or \textit{transcribed}, to make an RNA molecule. (In eukaryotes, the RNA polymerase makes a strand of mRNA from DNA).
    \item \textbf{Translation}: The sequence of the mRNA is decoded, or \textit{translated}, to specify an amino acid sequence (i.e. protein).
    \item \textbf{Gene}: A region of DNA (deoxyribonucleic acid) coding either for the messenger RNA encoding the amino acid sequence in a polypeptide chain or for a functional RNA molecule.
    \item \textbf{Chromosome}: Cellular structures that contain genes. A chromosome comprises of a single DNA molecule that may be either circular or linear.
    \item \textbf{Virus}: A small infectious agent that replicates only inside the living cells of other organisms.
    \item \textbf{Gene Expression}: The generation of a functional gene product from the information encoded by a gene, through the processes of transcription and translation. Gene products are often proteins. \footnote{Non-protein coding genes can encode functional RNA, including ribosomal RNA (rRNA), transfer RNA (tRNA) or small nuclear RNA (snRNA)}
\end{itemize}

\subsubsection{Central Dogma}
\begin{figure}[h]
\centering
\includegraphics[width=0.5\textwidth]{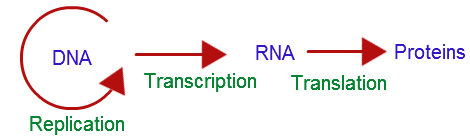}
\caption{Central dogma of Molecular Biology}
\end{figure}

The central dogma of biology states that:
DNA is \textit{transcribed} by RNA polymerases into mRNA (messenger RNA), which is read by ribosomes to generate protein through \textit{translation}. \cite{crick1970central}

In this project, our focus is on translation. More formally, we will be studying the translational interaction between the EV-A71 virus and its host cell, by observing gene expression values at different timepoints.

\subsubsection{Virus-Host Interaction}

When a host gets infected with a virus, the virus hijacks the host cell’s translation machinery, which affects protein production.

In the context of EV-A71, the viral replication process begins when a virus infects its host by attaching to the host cell and penetrating the cell wall or membrane. The viral genome then hijacks the host cell's translation mechanism, forcing it to replicate the viral genome by producing viral proteins to make new capsids (the protein shell of a virus that protects it). The viral particles are then assembled into virions. These virions burst out of the host cell during a process called lysis, killing the host cell. Some viruses take a portion of the host's membrane during the lysis process to form an envelope around the capsid. \cite{ScitableVirus}

After infection with any virus, overall translation of all genes decreases.
\begin{figure}[h]
    \centering
    \includegraphics[width=0.8\textwidth]{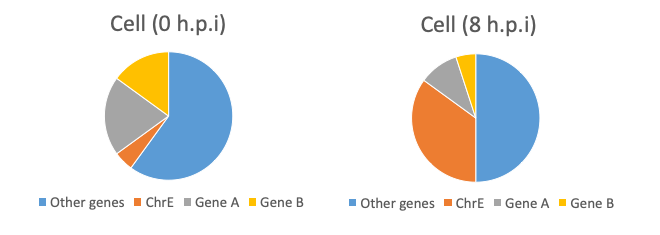}
    \caption{EV-A71 (chrE) infection over time*}
    \label{fig:virus}
\end{figure}

{\tiny *where Chromosome E (chrE) specifies the viral genes}

In Figure \ref{fig:virus},
Following viral replication, the new viruses may go on to infect new hosts. Many viruses such as influenza, chicken pox, AIDS, the common cold, and rabies cause diseases in humans.

\subsubsection{Hand, Foot, and Mouth Disease}

Hand, foot, and mouth disease (HFMD) is a common infection caused by a group of viruses. The one we will be examining in this report is the enterovirus A71, or the EV-A71 virus. HFMD mostly affects small children, occasionally causing symptoms to develop in adults. It typically begins with a fever, which is followed by rashes and bumps on other parts of the body. Currently, there is no treatment that specifically targets the disease, nor is there a vaccine that is approved for use in Singapore \cite{ang2009epidemiology}.

\begin{figure}[h]
    \centering
    \includegraphics[width=0.9\textwidth]{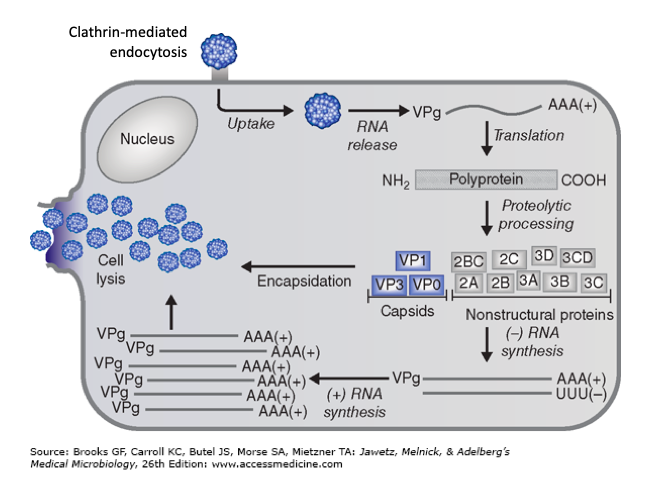}
    \caption{EV-A71 life cycle}
\end{figure}

\paragraph{GTPases}
\label{GTPAse}

Viruses depend on host cells for propagation and dissemination. Host cell mechanisms that allow viral entry, facilitate viral replication, and enable viral egress, are targeted for exploitation by viral pathogens \cite{amorim2018comprehensive}.

Guanosine-5'-triphosphate (GTP) is a purine nucleoside triphosphate. It is one of the building blocks needed for the synthesis of RNA during the transcription process, and is involved in energy transfer within the cell. GTPases are a class of proteins that bind and hydrolyze GTP. GTPases include Rab, Ras, Rac, Ran, Rho, Arf, and Sar proteins.

Our focus is on the proteins Rab and Rho. Emerging evidence demonstrates that viruses have evolved numerous strategies to modulate Rab proteins’ functions \cite{amorim2018comprehensive}.

In polioviruses, the 2C protein possesses weak GTPase activity. The 2C protein in EV-A71 viruses plays an important role in viral replication \cite{wang2018divergent}. 
Our hypothesis is that a certain member protein of GTPase is important for establishing the infection in EV-A71 viruses, and could be linked to Rho GTPases. We focus on the translation level changes in attempt to find information supporting our conjecture.

\subsubsection{Objective}

When a virus infects a cell, most of the processes in the cell get shut down and some proteins do not get generated. There are groups of:-
\begin{enumerate}
    \item Genes that escape this shut down by the virus \label{imp}
    \item Genes are affected by the shutdown
    \item Genes that are not influenced by the virus
\end{enumerate}

[\ref{imp}] is the most important, as these are the ones that the virus needs for its own production.

Thus, our objective is to identify these groups from the set of genes that we have available. We do this by using unsupervised learning techniques on gene expression data to generate clusters.

\subsection{Data Generation through Genome Sequencing Techniques}
\label{typesofdata}
RNA sequencing refers to techniques used to determine the sequence of RNA molecules. It includes high-throughput shotgun sequencing of cDNA molecules obtained by reverse transcription from RNA, and next-generation sequencing technologies to sequence the RNA molecules within a biological sample in an effort to determine the primary sequence and relative abundance of each RNA molecule.

For this project, we primarily use gene expression data that has been generated using the following techniques:-

\begin{itemize}
    \item \textbf{RNASeq}: RNASeq uses next-generation sequencing (NGS) to reveal the presence and total quantity of RNA in a biological sample at a given moment.
    \item \textbf{Ribosome Profiling (RPF)}: RPF uses specialised messenger RNA (mRNA) sequencing to determine which mRNAs are being actively translated. Unlike RNASeq, which sequences all of the mRNA of a given sequence present in a sample, RPF only targets mRNA sequences that are being actively translated.
\end{itemize}

\subsubsection{Methodology}
\label{experimentbio}

We posit that host factors regulated in viral translation may be important for establishing viral infection. The infection and analysis was carried out as follows -

\begin{enumerate}
    \item Infect cells at various timepoints
    \item Collect lysates post infection (cold synced) at various timepoints - 0 hours post infection (h.p.i), 2 h.p.i, 4 h.p.i, 6 h.p.i, 8 h.p.i
    \item Perform clustering analysis to identify relevant genes
    \item Validate changes in genes
    \item Test for functions of genes in viral infection
\end{enumerate}

% SEC2
\clearpage
\section{Literature Review}
\label{sec:LitReview}

\subsection{Clustering Time Series Gene Expression Values} \label{models}

Clustering techniques are essential in the data mining process, as they reveal natural structures and identify interesting patterns in the underlying data.

For gene expression data observed at different time-points, clustering can help identify biologically relevant groups of genes and samples.

Through the principle of guilt-by-association \cite{quackenbush2003microarrays}, if we find a known cluster that contains some uncharacteristic genes, we can hypothesise that these genes are significant.

As mentioned in \cite{d2005does}, there is no one-size-fits-all solution to clustering, or even a consensus of what an ideal clustering should be like. Clusters may have arbitrary shapes and sizes, and each clustering criterion imposes a certain structure on the data. If the data happen to conform to the requirements of a particular criterion, the true clusters are recovered; however, very rarely is this the case. Each algorithm imposes its own set of biases on the clusters it determines. Most (reasonable) clustering algorithms would yield similar results on synthetically constructed datasets; however, in practice, they can give widely differing results on real-world,noisy, gene expression data.

For temporal gene expression data, the most commonly used clustering algorithms \cite{friedman2000using,jiang2004cluster,schliep2003using} are
\begin{enumerate}
    \item Model Based Clustering
    \begin{itemize} 
        \item K-Means 
        \item Gaussian Mixture Models
        \item Hidden Markov Models
        \item Bayesian Networks
    \end{itemize}
    \item Hierarchical Clustering
    \item Self Organising Maps
\end{enumerate}

For all our model based clustering techniques, we determine partitions by using the expectation-maximisation (EM) algorithm (described in \ref{EM}) for maximum likelihood estimates of model parameters .

In this report, we will use hierarchical clustering for exploratory analysis, and apply all the model based clustering techniques to the different datasets we are provided with.

\subsection{Algorithms and Models}

\subsubsection{Expectation Maximisation (EM) Algorithm}
\label{EM}

The expectation maximisation (EM) algorithm is a general technique for finding maximum likelihood solutions for probabilistic models having latent variables \cite{bishop2006pattern}. EM is used for all the model based techniques covered in this report. 
\simpleparagraph{Procedure}

Consider a probabilistic model in which we collectively denote all the observed variables by $\mathbf{X}$ and the hidden variables by $\mathbf{Z}$. The joint distribution is governed by a set of parameters, denoted $\theta$. We try to maximise the likelihood function, that is given by 
\begin{equation}
    p(\mathbf{X}|\theta) = \sum_{\mathbf{Z}} p(\mathbf{X},\mathbf{Z} | \theta)
\end{equation}
If $\mathbf{Z}$ is continuous, we replace the summation with integration as appropriate.
The maximum likelihood estimator is 
\begin{equation*}
    \theta = \argmax_{\theta} f(\theta)
\end{equation*}
Direct optimisation of $p(\mathbf{X}|\theta)$ is difficult due to the presence of latent variables, so we attempt to optimise the expected value of the log-likelihood of the data given our model. We are now ready to define the general EM algorithm in algorithm \ref{alg: EMMain}.

\begin{algorithm}[h]
\caption{Expectation Maximisation}
\label{alg: EMMain}
\begin{algorithmic}
        \item \textbf{Initialisation:} For $l=0$, make an initial estimate of the parameters $\theta$.
        \item \textbf{E Step ($l^{th}$ iteration):} Evaluate  p($\mathbf{Z}$|$\mathbf{X}$,$\theta$), i.e find the conditional probability distribution of the latent variables given our current model $\theta^{(l)}$ and the data $\mathbf{X}$.
        \item \textbf{M Step ($l^{th}$ iteration):} Update $\theta$ as
        \begin{equation}
            \theta^{(l+1)} = \argmin_{\theta} u(\theta, \theta^{(l)}
        \end{equation}
        where 
        \begin{equation}
            u(\theta, \theta^{(l)}) = 
            - \sum_{\mathbf{Z}} p(\mathbf{Z}| \mathbf{X},\theta^{(l)}) \log p(\mathbf{X},\mathbf{Z}|\theta) 
            + \sum_{\mathbf{Z}} p(\mathbf{Z}| \mathbf{X},\theta^{(l)}) \log p(\mathbf{Z}|\mathbf{X},\theta^{(l)})
        \end{equation}
        \item \textbf{End:} Terminate when one of the following criteria is met
        \begin{enumerate}
            \item Iterate till $l$ reaches the maximum number of iterations specified
            \item Till the parameter estimates stop changing
            \item Till the likelihood function stop changing
        \end{enumerate}
        \end{algorithmic}

\end{algorithm}

\paragraph{Limitations}
Convergence of the EM algorithm is guaranteed by the Majorisation-Minimisation technique \cite{TanVYF}.
However, the EM Algorithm typically converges to the local optima. Thus, it needs to be run multiple times in order to find the best fitting model and its parameters.

\subsubsection{K-Means}

K-Means is an algorithm to identify groups, or clusters, of data points in a multidimensional space \cite{bishop2006pattern}. 

We group data points based on their feature similarities according to a given distance metric (Euclidean Distance in our case), by minimising a given cost function. 

Given a set of data points $D = \{x_t\}^n_{t=1}$ and fixed number of clusters $K \leq n$, the algorithm works as follows \cite{TanVYF}-

\begin{algorithm}[h]
\caption{K-Means Clustering}
\begin{algorithmic}
        \item \textbf{Step 0, Initialisation:} For $j=1,..,K$, initialise centers ${\mu_j}^{(1)}$ randomly.
        \item \textbf{Step $l \in \mathbb{N}$, Assignment (E) Step:} Assign each point $x_t$ to its closest mean
        \begin{equation*}
            {j_t}^{(l)} = \argmin_j {||x_t-\mu_j||}^2
        \end{equation*}
        \item \textbf{Step $l \in \mathbb{N}$, Update (M) Step:} Recompute $\mu_j$'s as means of the assigned points:
        \begin{equation*}
            \mu_j^{(l+1)} = \frac{1} { | \{ t : j_t^{(l)} =j\}|} \sum_{t : j_t^{(l)}=j} {x_t}
        \end{equation*}
        \item \textbf{End:} Terminate when cluster assignments $j_t$ do not change.
    \end{algorithmic}

\end{algorithm}

For a given partition $D = \cup_{j=1}^{K} D_j$,
where 
\begin{equation*}
    D_j:= \{ x \in D : j = \argmin_{j} ||x-\mu_j'||^2 \}
\end{equation*}
this algorithm minimises the overall squared error:
    \begin{equation}
        J(D_1,\dots,D_n,\mu_1,\dots,\mu_n)
        = \sum_{j=1}^K \sum_{x \in D_j}{{||x - \mu_j||}^2}
    \end{equation} \label{eq:k_means_cost}
where $D_1,D_2,\dots,D_n$ are the clusters and $\mu_1,\mu_2,\dots,\mu_n$ are cluster means. The two steps of updating the clusterings and re-estimating the centers correspond to the E step and M step of the EM Algorithm respectively.

Equation \eqref{eq:k_means_cost} is also known as the cost function.

The main problem of k-means is its dependency on the initially chosen centers. If the centers are chosen arbitrarily, then the algorithm converges very slowly or results in poor clustering. However, this can be solved by using the \textbf{K-Means++} \cite{arthur2007k} algorithm, which selects centers that are far apart from each other in a probabilistic manner. 

Through K-Means++, we choose the initial $K$ centers as follows \cite{TanVYF}-

\begin{enumerate}
    \item Take a random data point $\mu_1$ chosen uniformly from dataset $D$.
    \item Calculate new center $\mu_i$, by choosing $x \in D$ with probability $\frac{D(x)^2}{\sum_{x' \in D} D(x')^2}$, (where $D(x)$ denotes the shortest distance from the data point $x$ to its closest, already chosen center)
    \item Repeat the previous step till we have $K$ centers.
\end{enumerate}

\paragraph{Limitations}
Due to its ease of implementation and termination in a finite number of steps, K-Means is one of the most popular (and usually one of the first) algorithms applied for unsupervised learning problems. 

However, due to the nature of the distance function chosen (Euclidean), K-means doesn't capture temporal dependencies in the data. Thus, it is not entirely suitable for our time series dataset. Moreover, k-means does hard clustering, i.e. a point can belong to only one cluster, while in real life data, a point can belong to more than one cluster.

\subsubsection{Gaussian Mixture Models}

Gaussian mixture models (GMMs) are mixtures, or linear combinations, of models with Gaussian distributions. They are widely used in data mining, statistical analysis and pattern recognition problems \cite{bishop2006pattern}. 

\begin{figure}[ht]
    \centering
    \includegraphics[width=0.5\textwidth]{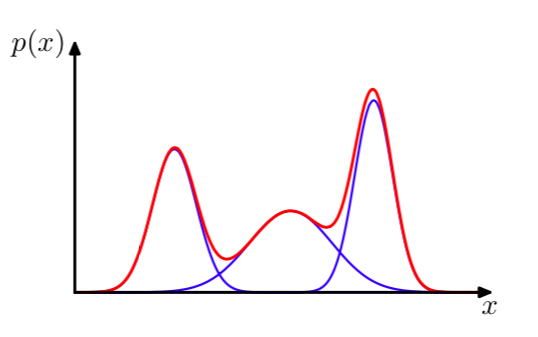}
    \caption{Mixture of 3 Gaussians in one dimension}
    \label{fig:GMM}
\end{figure}

A Gaussian mixture model is represented by 
\begin{equation} \label{eq: GMM}
    p(\mathbf{x)} = \sum_{k=1}^{K} \pi_k \mathcal{N} ( \mathbf{x} | \mu_k, \Sigma_k)
\end{equation}

where the equation above represents a mixture of $K$ Gaussians, $\mathcal{N} ( \mathbf{x} | \mu_k, \Sigma_k)$ is a \textit{component} of the mixture model (each having its own mean $\mu_k$ and covariance $\Sigma_k$, and $\pi_k$ are the component weights, such that $\sum_k \pi_k = 1$.

To use GMMs for clustering, we assume that our data are generated i.i.d from a mixture of $K$ Gaussians, and we apply EM to estimate the $K$ means, covariances, and weights, and thus determine the clusterings. We achieve this by introducing a latent variable $z$ in \eqref{eq: GMM}. This latent variable has a 1-of-K representation - thus $z_k \in \{0,1\}$ and $\sum_k z_k=1$ (for each $x$), and find $p(\mathbf{z})$ and $p(\mathbf{x|z})$, to get
\begin{equation} \label{eq: GMM_latent}
    p(\mathbf{x)} = \sum_z p(\mathbf{z})p(\mathbf{x|z})= \sum_{k=1}^{K} \pi_k \mathcal{N} ( \mathbf{x} | \mu_k, \Sigma_k)
\end{equation}
where $p(\mathbf{z}) = \displaystyle \prod_{k=1}^{K} \pi_k^{z_k}$ and $p(\mathbf{x|z}) = \displaystyle \prod_{k=1}^{K} \pi_k^{z_k} \mathcal{N} ( \mathbf{x} | \mu_k, \Sigma_k)^{z_k} $. 

Another quantity that is important is the conditional probability of $\mathbf{z}$ given $\mathbf{x}$, also known as the \textit{responsibility} that component $k$ takes for producing the observation $\mathbf{x}$, denoted by 

\begin{equation} \label{eq:respons}
    \gamma(z_k) = p(z_k = 1|x) = \frac{\pi_k \mathcal{N}(x|\mu_k,\Sigma_k)}{\sum_{j=1}^{K} \pi_j \mathcal{N} (x|\mu_j, \Sigma_j)}
\end{equation}

These formulas containing latent variable $z$ help us formulate EM for GMM, given in \ref{alg:EMGMM}.

\begin{algorithm} [h]
\caption{EM for Gaussian Mixture Models}
\label{alg:EMGMM}
\begin{algorithmic}
            \item \textbf{Init}: Initialise the means and covariances $\mu_k$ and $\Sigma_k$ and mixing co-efficients $\pi_k$, and evaluate the initial value of the log-likelihood.
        \item \textbf{Step l, E step}: Evaluate the responsibilities from \eqref{eq:respons} using the current parameter values.
        \item \textbf{Step l, M Step}: Re-estimate the parameters using the current responsibilities.
        \begin{equation*}
             \mu_k^{(l+1)} = \frac{1}{N_k} \sum_{n=1}^{N} \gamma(z_{n,k}) \mathbf{x}_n 
             \end{equation*}
             \begin{equation*}
                             \Sigma_k^{(l+1)} = \frac{1}{N_k} \sum_{n=1}^{N} \gamma(z_{n,k}) (\mathbf{x}_n - \mu_k^{(l+1)})(\mathbf{x}_n - \mu_k^{(l+1)})^T
        \end{equation*}
        \begin{equation*}
            \pi_k^{(l+1)} = \frac{N_k}{N}
        \end{equation*}
        where $N_k = \sum_{n=1}^N \gamma(z_{nk})$
        \item Evaluate the log-likelihood
        \begin{equation*}
            \log p(X|\mu,\Sigma, \pi) = \sum_{n=1}^N \log \{ \sum_{k=1}^K \pi_k \mathcal{N} (x_n|\mu_k,\Sigma_k) \}
        \end{equation*}
        and \textbf{terminate} if one of the following criteria are met- 
        \begin{enumerate}
            \item Paramater estimates have converged (stopped changing).
            \item Log likelihood has converged.
            \item Maximum umber of iterations $l$ have been reached.
        \end{enumerate}
        otherwise return to Step 2.
        \end{algorithmic}
\end{algorithm}

The clusterings are indicated by the latent variable $\mathbf{z}$.

\begin{figure}[ht]
    \centering
    \includegraphics[width = 0.45\textwidth]{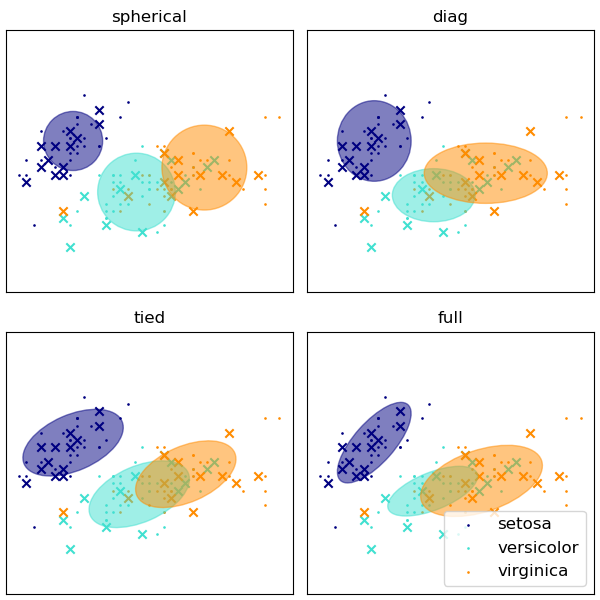}
    \caption{Different covariance settings for the 2D Iris dataset}
    \label{fig:Covariance}
\end{figure}

\newpage
\paragraph{Covariance Matrix}
Having different settings of the covariance matrix for each component help in restricting the number of parameters that need to be estimated (thus optimising the runtime of the algorithm). However, some of the settings (diagonal, tied and spherical) restrict the contour of the clusters obtained, as seen in figure \ref{fig:Covariance}. In the diagonal setting, the elliptical clusters obtained are aligned with the co-ordinate axes, while the spherical clusters are circular, while all clusters have the same shape in the tied setting. Assuming we have $k$ components, here are 4 different covariance settings, summarised in the table \ref{tab:GMM_Cov}. 
\begin{table}[h]
    \centering
    \begin{tabular}{|c|c|c|}
        \hline
         \textbf{Covariance} & \textbf{Description} & \textbf{Parameters} \\    \hline

         Full & \pbox{20cm}{\hspace{0.1cm}\\Each component has its own \\ general covariance matrix\\} & \small $\frac{KD(D+1)}{2}$ \\     \hline

         Diagonal& \pbox{20cm}{\hspace{0.1cm}\\Each component has a\\ diagonal covariance matrix\\} & $KD$\\    \hline

         Spherical & \pbox{20cm}{\hspace{0.1cm}\\Each component \\ has a single variance\\} & $K$\\    \hline

         Tied & \pbox{20cm}{\hspace{0.1cm}\\All components share the\\ same general covariance matrix\\} &  $\frac{D(D+1)}{2}$\\   \hline

    \end{tabular}
    \caption{Covariance Settings for GMM}
    \label{tab:GMM_Cov}
\end{table}

\paragraph{Limitations}
As discussed for previous models, GMM converges to a local optima, so we need to run the model multiple times to determine the best fit. Moreover, GMM takes a long time to run - however, we can indicate different parameter settings for the covariance matrix, based on what we know (or assume) about the data. This helps keeps the number of parameters that we are estimating under control and reduces the computation time.

\subsubsection{Hidden Markov Models}
In a first-order Markov chain describing a sequence of possible events, the probability of each event depends only on the state attained in the previous event.
\begin{figure}[ht]
    \centering
    \includegraphics[width=0.4\textwidth]{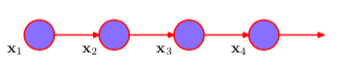}
    \caption{First-order Markov Chain}
    \label{fig:FOMC}
\end{figure}

A hidden Markov model can be viewed as a Markov chain of latent variables, with each observation conditioned on the state of the corresponding latent variable. The states of the Markov chain are not observed directly, i.e. the Markov chain itself remains hidden \cite{bishop2006pattern}.

\begin{figure}[ht]
    \centering
    \includegraphics[width=0.7\textwidth]{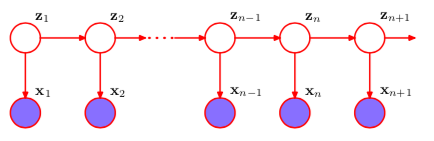}
    \caption{Hidden Markov Model}
    \label{fig:my_label}
\end{figure}

A single time slice of HMM corresponds to a mixture distribution, with component densities given by $p(x|z)$. Therefore HMMs can also be interpreted as an extension of a mixture model in which the choice of mixture component for each observation is not selected independently but depends on the choice of component for the previous observation \cite{bishop2006pattern}.

HMMs are widely used in speech recognition, natural language modelling, on-line handwriting recognition and for the analysis of biological sequences such as proteins and DNA \cite{bishop2006pattern}.
Thus, we can use a Hidden Markov Model (HMM) to model temporal dependencies in the data. 

An HMM contains 3 essential components - observed sequence of data $x_i$, hidden states $z_i$, transition probabilities $p(z_i|z_{i-1})$ and emission probabilities $p(x_i|z_i)$.

The hidden states in an HMM are analogous to latent variables in GMMs, responsible for generating the given observations. The probability distribution of a hidden state $z_n$ depends upon the state of the previous hidden state $z_{n-1}$. This is encoded by a transition probability matrix. For an HMM with $M$ hidden states, the transition probability matrix has $M(M-1)$ independent parameters.

The emission probabilities are conditional distributions of the observed variables are $p(x_n|z_n, \phi)$, where $\phi$ is a set of parameters governing the distribution. The $\phi$ may be given by Gaussians if the elements of $x$ are continuous variables, or by conditional probability tables if $x$ is discrete \cite{bishop2006pattern, glass_zue_2003}.

There are 3 basic problems tackled by hidden Markov models \cite{TanVYF,glass_zue_2003} - 
\begin{enumerate}
    \item Given an observation sequence $\mathbf{x}$ and a model $\lambda$, compute the probability $p(x|\lambda)$. This is achieved using the Forward-Backward algorithm.
    \item Given an observation sequence $\mathbf{x}$, compute the most likely observed state sequence. This is done using the Viterbi algorithm.
    \item Adjust the model parameters of $\lambda$ to maximise $p(x|\lambda)$ using the Baum-Welch algorithm.
\end{enumerate}

\paragraph{HMM-based Clustering}
We can apply the EM algorithm defined in Algorithm \ref{alg: EMMain} to do clustering using HMMs.

The EM procedure for HMM based clustering is fairly complex, as we need to use a sub-EM procedure (Baum-Welch algorithm) in the M step, $K$ times. 

Starting from an initial collection of HMMs, EM iteratively finds cluster assignments that maximise the joint likelihood of the clustering.

\textbf{Initialisation:} Given sequences $O^i$, where $i$ $\in$ [1,$n$], we randomly generate $K$ HMMs $\lambda_1, \lambda_2,...,\lambda_k$ and a partitioning $C = {C_1, C_2,...C_k}$, where the assignment to clusters is done to maximise the following objective function:
\begin{equation}
   f(C) = \prod_{k = 1}^{K} \prod_{i \in C_k} L(O^i|\lambda_k) 
\end{equation}

where $L(O^i|\lambda_k)$ defines the likelihood function, i.e. the probability density for generating sequence $O^i$ by model $\lambda_k$.

Assuming that we have a collection of $K$ HMMs, our algorithm is as follows:

\begin{enumerate}
    \item \textbf{Iteration} ($t$ $\in$ \{1,2,...\}):
    \begin{enumerate}
        \item \textbf{E Step:} Generate a new partitioning of the sequences by assigning each sequence $O^i$ to the model $k$ for which the likelihood $L(O^i|\lambda_k)$ is maximal.
        \item \textbf{M Step:} Re-estimate parameters using the Baum-Welch algorithm \cite{glass_zue_2003}: ${\lambda_1}^t,{\lambda_2}^t...{\lambda_k}^t$ using their start parameters from the previous iteration and the sequences assigned to each model.
    \end{enumerate}
    \item \textbf{Stop}, if one of the following takes place -
    \begin{enumerate}\label{Criterion}
        \item If the improvement of the objective function is below a given threshold
        \item There are no more re-assignments
        \item The number of iterations is reached
    \end{enumerate}
\end{enumerate}

\paragraph{Limitations}
The biggest limitation with HMM-based clustering is convergence to a local optima, so the algorithm needs to be run multiple times with different random initialisations to find the best fit. However, like GMM training, HMM based clustering takes significant computation time - especially due to the Baum-Welch EM procedure in the M step of the clustering.

\subsubsection{Bayesian Networks}

Bayesian networks are probabilistic graphical models that represent a set of variables and their conditional dependencies via directed acyclic graphs.

In \cite{TanVYF}, we studied the Chow-Liu algorithm to learn  tree-structured graphical models. In \cite{pham2009unsupervised}, the authors combined Chow-Liu Trees and Expectation Maximisation to derive a model-based clustering method, called the CL Multinet Classifier.

If there are $K$ classes, then $K$ CL trees are learned. Each tree distribution $Pr_k$ approximates the joint probability distribution of the attributes, given a specific class. The root attribute has no parents, while the other attributes can have at most ONE parent, i.e. if the attributes are expressed as r.vs $\{X_1, X_2,...,X_n\}$, then $\Pi_{X_i} = \{X_{j \neq i}\}$ has only one attribute, and $\Pi_{X_{root}} = \{\emptyset\}$. 

An inherent advantage of applying Bayesian networks case is that we can infer the structure behind each cluster too. For each cluster, we can see which timepoints affect the other. 
 
Using Bayes Theorem, the posterior probability of a Chow Liu Tree can be expressed as
 \begin{equation}
         Pr(C=k|X_1,X_2,...,X_n) =\frac{ Pr(C=k)Pr(X_1,X_2,...,X_n|C=k)}{\sum_{k'}Pr(C=k')Pr(X_1,X_2,...,X_n | C = k')}
 \end{equation}
We observe that the denominator is constant with respect to the class, so we express it as $\frac{1}{\beta}$. 
 
Then, the previous formula can be re-written as
 
\begin{equation}
    Pr(C=k|X_1,X_2,...,X_n) = \beta*Pr(C=k)\prod_{i=1}^{n} Pr_k(X_i | \Pi_{X_i})
\end{equation}
 
The edge weights of a Chow Liu tree are encoded by the \textit{mutual information}, which is derived from the empirical probability distributions observed from the data.
 
 \begin{equation}
     I(X,Y) = \sum_{x \in X} \sum_{y \in Y} Pr(x,y) \log \frac{Pr(x,y)}{Pr(X)Pr(y)}
 \end{equation}
 
 A maximum weighted spanning tree (either Prim's/Kruskal's) algorithm is then used to find the Chow-Liu tree for each $k^{th}$ class.
 
 Unsupervised training of the CL Trees (henceforth referred to as CL Multinets) is carried out assuming that the data have been generated from a mixture of $K$ Bayesian networks. This mixture can be described as 
 \begin{equation}
     Pr(X) = \sum_{k=1}^{K} \alpha_k f_k(X)
 \end{equation}
where 
\begin{equation}
    \sum_{k=1}^{K} \alpha_k = 1, \hspace{0.5cm}\alpha_k \geq 0
\end{equation}

The $\alpha_k$ are the mixing coefficients and the $f_k$ are the joint probability distributions of the CL Multinets.  Given $N$ observations, $K$ clusters (CL Multinets), and initial partitions P$^0$=\{P$_0^0$,P$_1^0$,...,P$_K^0$\}, the EM procedure for the $m^{th}$ iteration is detailed in algorithm \ref{alg: EMBN}.

\begin{algorithm}[h]
\small
\caption{Expectation Maximisation for Bayesian Networks}
\label{alg: EMBN}
\begin{algorithmic}
     \item \textbf{E-Step:} For $r=1,2,...,N$ and $k=1,2,...,K$ compute the posterior probabilities for $x^r$ belonging to P$_k$.
     \begin{equation}
         t_k^m(x') = \frac{\alpha_k^m \prod_{i=1}^{n} Pr_k^m(x_i^r|\Pi_{x_i^r})}{\sum_{k'=1}^{K}\alpha_{k'}^m \prod_{i=1}^{n} Pr_{k'}^m(x_i^r | \Pi_{x_i^r})}
     \end{equation}
     \item \textbf{C-Step:} Update the partition $P^m$=\{P$_0^m$,P$_1^m$,...,P$_K^m$\} by assigning each $x^r$ to the cluster that provides the max posterior probability.
     \item \textbf{M-Step:} For $k=1,...,K$, maximise the Classification Maximum Likelihood (CML) criteria and re-estimate the parameters $\theta^m$ using the new partitions $P_k^m$.
     \begin{equation} \label{eq:CML}
         CML = \sum_{k=1}^{K}\sum_{x^r \in P_k} \log \prod_{i=1}^{n} Pr_k(x_k^r | \Pi_{x_i^r}) + \sum_{k=1}^{K} n_k \log \alpha_k
     \end{equation}
      where $n_k$ are the number of samples belonging to the $k^{th}$ cluster, and $\alpha_k$ are the cluster weights.
      
      The parameters for the second term are re-estimated (maximised) using the following formula:
      \begin{equation}
          \alpha_k^{m+1} = \frac{n_k}{N}, \hspace{0.5cm} k=1,...,K
      \end{equation}
       The first term (after some manipulation) contains the mutual information computed by the observed data (empirical distributions). Using the maximum weighted spanning tree algorithm, we maximise \eqref{eq:CML}, resulting in a new tree distribution.
       \item \textbf{Terminate:} If any of the following criteria are reached:
       \begin{enumerate}
           \item Maximum number of iterations reached.
           \item CML converges.
           \item Parameters converge.
       \end{enumerate}
\end{algorithmic}
\end{algorithm}
\normalsize
\newpage
\paragraph{Limitations}
Although Bayesian Networks are directed acyclic graphs and thus use fewer parameters than HMMs, unlike HMMs, we need to discretise the data in order to use the empirical frequencies to estimate joint probability distributions. Discretisation also results in loss of information, as the resultant dataset is dependent upon the number of bins chosen. Moreover, just like HMM-based clustering, there exist only general information theoretic criteria to determine the number of clusters $K$, which is not always be applicable \cite{pham2009unsupervised}.

\subsubsection{Hierarchical Clustering}

All the previous models that we covered can be classified under \textit{partitional clustering}. The other category, hierarchical clustering, seeks to build a hierarchy of clusters using a given distance metric for datum and a similarity criteria for the clusters obtained. However, unlike partitional clustering, there exist no definitive statistical techniques determine the optimal number of clusters - instead, we need to examine the resulting hierarchical structure (dendogram) and visually determine the optimal number of clusters \cite{maimon2005data}.
 
 There are two approaches to hierarchical clustering - 
 \begin{itemize}
     \item Agglomerative: Each datum initially represents a cluster of its own. Clusters are then successively merged until the desired cluster structure is obtained.
     \item Divisive: All objects initially belong to one cluster. This main cluster is then divided into sub-clusters, which are successively sub-divided into their own sub-clusters, and this process continues until the desired cluster structure is obtained.
 \end{itemize}

We set the distance metric (for within cluster distances) as Euclidean. 
The linkage criterion in hierarchical clustering specifies the difference between any two sets (clusters) obtained. It determines which clusters to merge by calculating the differences between two clusters as a function of the pairwise distances between observations. The different types of linkages are-
\begin{itemize}
    \item Single: the distance between two clusters is equal to the shortest distance from any member of one cluster to any member of the other cluster. 
    \item Average: the distance between two clusters is equal to the average distance from any member of one cluster to any member of the other cluster.
    \item Complete: the distance between two clusters is equal to the longest distance from any member of one cluster to any member of the other cluster.
    \item Ward: the distance is defined as the error function of the unified cluster minus the error functions of the individual clusters, where the error function is the average distance of each data point in a cluster to the cluster centroid.
\end{itemize}

\begin{figure}
    \centering
    \includegraphics[width=0.6\textwidth]{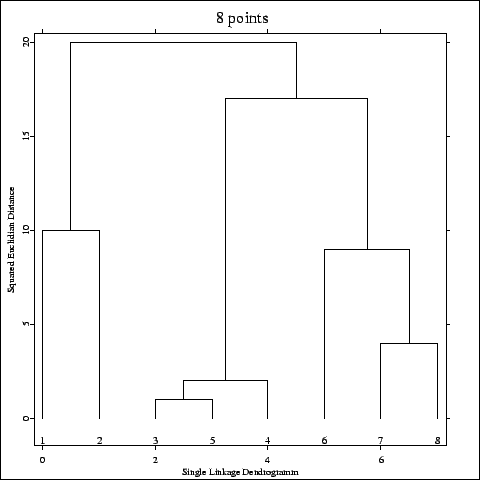}
    \caption{Hierarchical Clustering Dendogram for Single-Linkage}
    \label{fig:HC}
\end{figure}

For this report, we use agglomerative hierarchical clustering \textit{not} in model building, but in Section \ref{FeatureEngineering} for feature engineering , so as to obtain a rough estimate of the number clusters we may expect from the data.

\subsection{Evaluation Metrics for Model Selection} \label{Eval}

Given data, model selection is the task of selecting the best statistical model from a set of candidate models. 

In our clustering analysis, the main conundrum is determining the number of clusters.

The estimation of the true number of classes has been recognised as “one of the most difficult problems in cluster analysis" \cite{bock1996probabilistic}. The correct choice of $K$ is often ambiguous - sometimes, choosing the best fitting model is not the wisest choice. Consider increasing $K$ without any penalty - this will keep reducing the error but the model is less than ideal (e.g. In K-Means, if we have $n$ clusters for $n$ points, each point is its own center, reducing the cost function to 0). 

Moreover, given multiple candidate models having similar accuracy, the simplest model is most likely to be the best choice (Occam's razor).

Thus, the optimal choice of $K$ balances accuracy and number of parameters required for the model. This is tackled in \ref{conundrum1} and \ref{con1}.

A secondary conundrum is comparing two different clustering models, given that we obtain "ground truth" values for some of the clusters. This is handled in \ref{condundrum2}. 

\subsubsection{Information Criteria} \label{conundrum1}

When a statistical model is used to represent the process that generated the data, the representation will almost never be exact. Some information is thus lost by the model. Bayesian Information Criteria (BIC) and Akaike Information Criteria (AIC) \cite{burnham2004multimodel} deal with estimating this lost information, and calculate the trade-off between the \textit{goodness of fit} of the model and the \textit{complexity} of the model.

\paragraph{BIC and AIC}

BIC and AIC penalise the total number of parameters. Both are functions of the maximised log-likelihood of the model and the estimated number of parameters - however, BIC penalises model complexity more heavily. We define the AIC and BIC as follows-

\begin{equation*}
    AIC =  \log L(\theta) - P \\
\end{equation*}

\begin{equation*}
        BIC =  \log L(\theta) - \frac{P}{2}\log(N) 
\end{equation*}

where $L(\theta)$ is the maximised log-likelihood of the model $\theta$, $P$ is the number of free parameters and $N$ is the number of data points. For a range of $K$ values, we calculate AIC/BIC, and choose the $K$ for which this criterion is \textit{maximum}. Thus, higher AIC or BIC values indicate better fitting models.

\subsubsection{Elbow Method}\label{con1}
\begin{figure}[ht]
    \centering
    \includegraphics[width=0.7\textwidth]{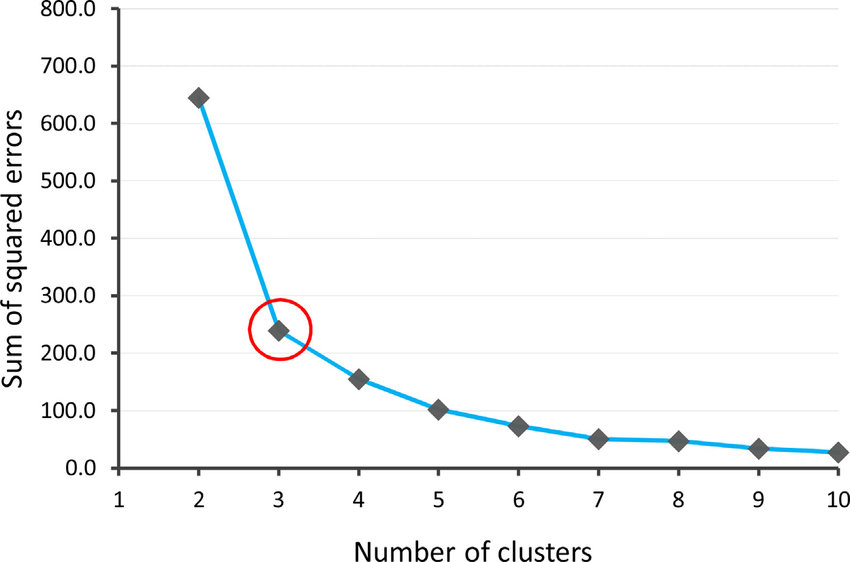}
    \caption{Elbow method for k-means clustering}
    \label{fig:elbow}
\end{figure}

The elbow method \cite{kodinariya2013review} is a popular method to determine the true number of clusters by plotting the cost function for a given clustering algorithm against different $K$, and observing the point where the graph makes a sharp \textit{elbow} shape, as seen in Figure \ref{fig:elbow}. More formally, this method looks at the percentage of variance explained as a function of the number of clusters. In essence, the elbow method observes the point of sharpest decrease in the cost function, i.e. the most improvement to the cost function, and estimates that to be the ideal number of clusters.
However, the limitation of this method is that the elbow point cannot always be unambiguously identified - sometimes there is no elbow, or sometimes there are several elbows.

\subsubsection{Misclassification Error (ME) Distance}\label{condundrum2}

ME distance is used to calculate the similarity between two different clusterings of the same data. Given the "ground truth" (i.e. true classes of each point), we can compare the clustering obtained by our technique with the ground truth to see how our model compares. ME distance thus represents the well known cost of classification, minimised over all permutations of the labels \cite{meila2005comparing}. 

It is calculated as follows -

Consider two clusterings $\mathbb{C} =\{C_1,C_2,...,C_K\}$ and $\mathbb{C'} =\{C'_1,C'_2,...,C'_{K'}\}$. 
We define the confusion matrix of $\mathbb{C}$ and $\mathbb{C'}$ as a $K x K'$ matrix $M=[m_{kk'}]$, with $m_{kk'} = |\mathbb{C_k} \cap \mathbb{C'_{K'}}|$. We consider the case where both models cluster $N$ points, and $K=K'$. 

Then the ME distance is defined as
\begin{equation*}
    d(C,C') = 1 - \frac{1}{N} \max_{\pi \in \Pi_k} \sum_{k=1}^{K} {m_{k,\pi(k)}}
\end{equation*}
where $\Pi_k$ contains all permutations of the label $K$.

We want the ME Distance between our models and the ground truth to be close to 0.

\subsubsection{Gene Ontology}
Gene Ontology (GO) is a framework for the model of biology, unifying the representation of gene and gene product attributes across all species. We define concepts and classes used to describe gene functions, and observe the relationships between them.

The ontology covers 4 categories-
\begin{enumerate}
    \item Biological Process: operations or sets of molecular events with a defined beginning and end, pertinent to the functioning of integrated living units: cells, tissues, organs, and organisms.
    \item Cellular Component: the parts of a cell or its extracellular environment.
    \item Molecular Function: the elemental activities of a gene product at the molecular level.
    \item KEGG Pathway: pathways representing molecular interaction, reaction and relation networks for various biological processes.
\end{enumerate}

We use \textbf{DAVID} \cite{huang2007david} to perform gene ontology analysis on the clusters we generate.

Each cluster obtained from our model(s) contains genes that share similar biological properties. We perform GO analysis on all these different clusters, and attempt to identify terms that are related to infection - like viral process, translation, DNA repair, etc. If we do find terms related to viral process, we use guilt-by-association to find other significant terms.

\begin{figure}[ht]
    \centering
    \includegraphics[width=0.8\textwidth]{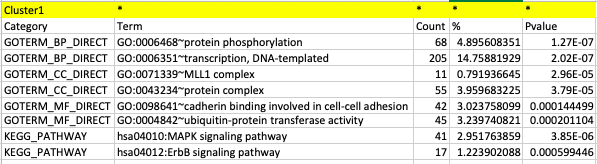}
    \caption{Sample GO for a cluster}
    \label{fig:GO}
\end{figure}

\label{PValue}

\textit{p-value} is the probability of seeing at least x number of genes out of the total n genes in the list annotated to a particular GO term, given the proportion of genes in the whole genome that are annotated to that GO term. In essence, the GO terms shared by the genes in the user’s list are compared to the background distribution of the annotation. Thus, the smaller the p-value is, the more significant the particular GO term associated with the group of genes is (i.e. the less likely the observed annotation of the particular GO term to a group of genes occurs by chance) \cite{huang2007david}.

% SEC 3
\clearpage
\section{Exploratory Analysis}
\label{sec:ExpAnal}

\subsection{Data Description}

We have 3 datasets consisting of unlabelled human gene expression values, obtained after infection with the EV-A71 virus.

\begin{itemize}
    \item \textbf{RNASeq}: 12,340 samples of RNASeq data at 5 timepoints
    \item \textbf{RPF}: 11,746 samples of RPF data at 5 timepoints.
    \item \textbf{TE}: 11,633 samples of TE data at 5 timepoints.
\end{itemize}

The attributes are gene expression values $x$ hours post infection (h.p.i), where $x \in \{0,2,4,6,8\}$. 

As described in Section \ref{typesofdata}, RNASeq represents the total amount of genes within the cell, while RPF represents the amount of genes being translated. TE measures the rate at which genes get translated.

\subsubsection{Data Preprocessing}
\begin{figure}[h]
    \centering
    \includegraphics[width=1.0\textwidth]{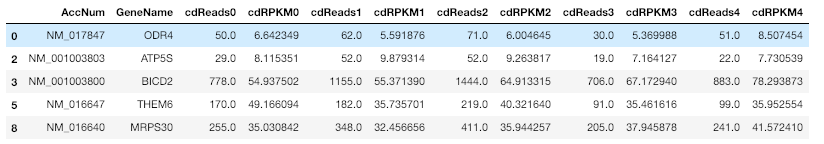}
    \caption{RNASeq Data}
    \label{fig:RNASeq}
\end{figure}

\textbf{cdRPKM} values are raw gene expression values; \textbf{cdReads} are the number of times a gene was read (as per the biological technique used), while the \textbf{AccNum} and \textbf{GeneName} represent how to ID the gene.

We first filter the dataset for cdReads $\geq 10$ to remove noise; this removes genes with very small reads, likely generated due to experimental errors.

Then, we transform the cdRPKM values to $\log_2$, which is a standard practice in bioinformatics.

\begin{figure}[h]
    \centering
    \includegraphics[width=1.0\textwidth]{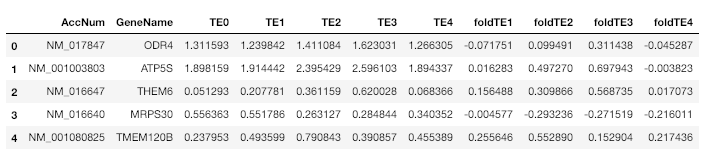}
    \caption{TE Data}
    \label{fig:TE}
\end{figure}

We can calculate TE data in Figure \ref{fig:TE} by dividing the raw RPF value by the raw RNASeq value, and transforming it to $log_2$. Alternatively, we can subtract the $log_2$ RNASeq value from the $log_2$ RPF value to also obtain TE.

\subsection{Initial Analysis}

\begin{figure}[h]
    \centering
    \begin{subfigure}[b]{0.3\textwidth}
        \includegraphics[width=\textwidth]{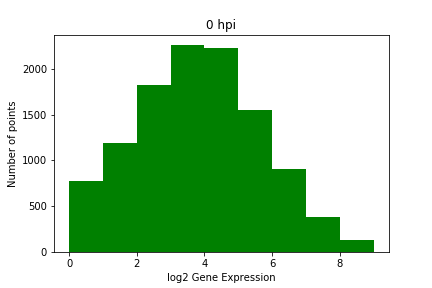}
        \caption{0 h.p.i}
    \end{subfigure}
    \begin{subfigure}[b]{0.3\textwidth}
        \includegraphics[width=\textwidth]{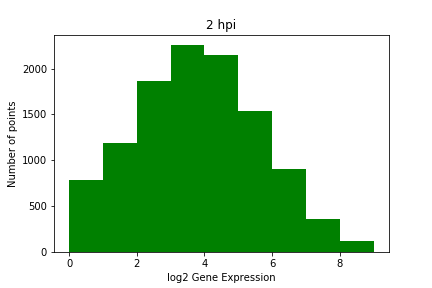}
        \caption{2 h.p.i}
    \end{subfigure}
    \begin{subfigure}[b]{0.3\textwidth}
        \includegraphics[width=\textwidth]{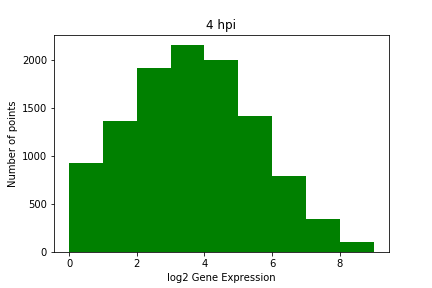}
        \caption{4 h.p.i}
    \end{subfigure}
    \begin{subfigure}[b]{0.3\textwidth}
        \includegraphics[width=\textwidth]{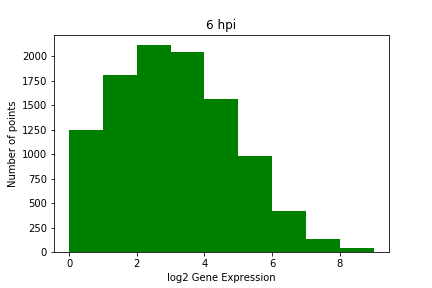}
        \caption{6 h.p.i}
    \end{subfigure}
        \begin{subfigure}[b]{0.3\textwidth}
        \includegraphics[width=\textwidth]{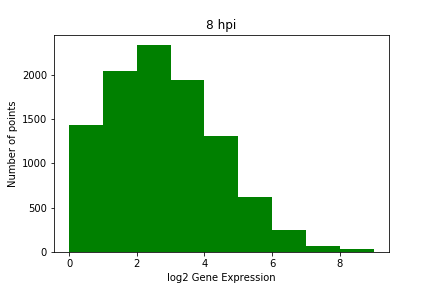}
        \caption{8 h.p.i}
    \end{subfigure}
    \caption{Distribution of $log_2$ values of RNASeq data}
    \label{fig:explPlotsRNA}
\end{figure}

\begin{figure}[h]
    \centering
    \begin{subfigure}[b]{0.3\textwidth}
        \includegraphics[width=\textwidth]{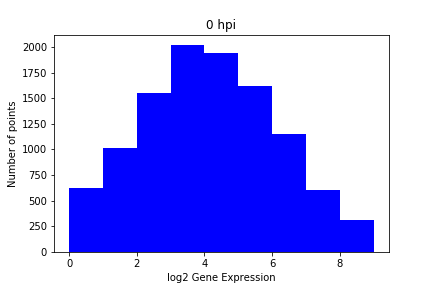}
        \caption{0 h.p.i}
    \end{subfigure}
    \begin{subfigure}[b]{0.3\textwidth}
        \includegraphics[width=\textwidth]{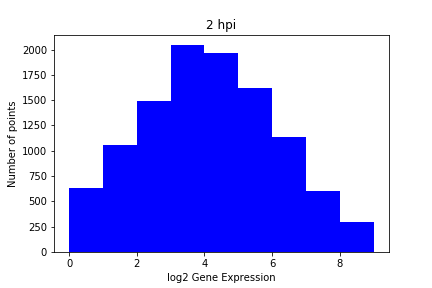}
        \caption{2 h.p.i}
    \end{subfigure}
    \begin{subfigure}[b]{0.3\textwidth}
        \includegraphics[width=\textwidth]{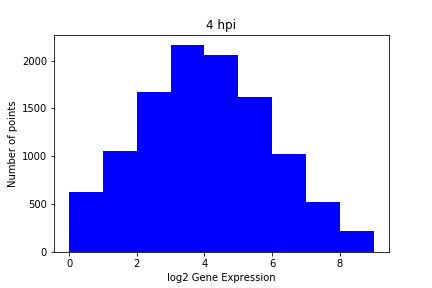}
        \caption{4 h.p.i}
    \end{subfigure}
    \begin{subfigure}[b]{0.3\textwidth}
        \includegraphics[width=\textwidth]{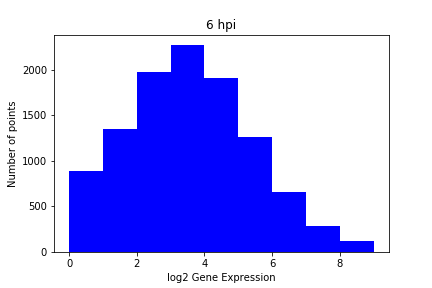}
        \caption{6 h.p.i}
    \end{subfigure}
        \begin{subfigure}[b]{0.3\textwidth}
        \includegraphics[width=\textwidth]{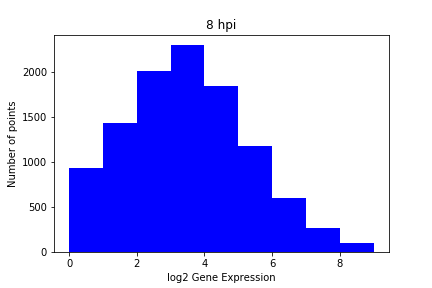}
        \caption{8 h.p.i}
    \end{subfigure}
    \caption{Distribution of $log_2$ values of RPF data}
    \label{fig:explPlotsRPF}
\end{figure}

For the RNASeq and RPF datasets, we plot individual timepoint $log_2$ values in Figures \ref{fig:explPlotsRNA} and \ref{fig:explPlotsRPF}.

From the plots, we infer that all timepoints follow a Gaussian distribution. we posit that Gaussian Mixture Models and Hidden Markov Models with continuous emissions will be useful in clustering our time-series gene expression values. 

For both datasets, we observe that the median keeps shifting to the left for later timepoints. This is due to the infection spreading- with the passage of time, more viral translation leads to an overall decrease in gene expression.

To standardise the medians for all timepoints, we again follow a standard practice in bioinformatics - we median-normalise each timepoint (subtract the median from every value) so that each timepoint has the same median (0).  

After median-normalisation, the distribution of the maximum and minimum values are generally similar for all time points.

% % SEC 4
% \input{./Sections/4_EvaluationMetrics}

% SEC 6
\clearpage
\section{Results}
\label{sec:Results}

We apply the models from Section \ref{models} to the datasets described in Section \ref{sec:ExpAnal}, and attempt to identify those groups of genes that are critical for viral translation. 

There are two tasks we need to perform - firstly, identify an optimal $K$, and secondly, decide the benchmark algorithm (the most optimal model) that we will compare all our other techniques with.

We evaluate the performance of the models through the methods described in Section \ref{Eval} and interpret the significance of the results obtained. 

We perform our analysis in a systematic way - 
\begin{enumerate}
\label{Analysis}
    \item A priori, we know that the datasets will not contain more than 20-30 clusters. We restrict our analysis for $2 \leq K \leq 30$.
    \item We apply model-based clustering techniques on the complete datasets, and use the evaluation criteria to determine the optimal $K$.
    \item If the results are inconclusive, we implement feature engineering in Section \ref{FeatureEngineering} and re-run the model-based clustering techniques on this modified data.
    \item Attempt to determine $K$ through BIC/AIC. 
    \item If no optimal value is found, do clustering for specific $K$ (as determined by biologists, as in Figure \ref{fig:IdealClustering}).
    \item Interpret results through GO analysis and/or other analyses.
\end{enumerate}

\subsection{Feature Engineering} \label{FeatureEngineering}

Since this is time-series data, we also quantify the change between successive timepoints.

From the RNASeq and RPF datasets, we find the amount of change between timepoints and add these as features to the original dataset. Formally known as $\log_2$ fold change in bioinformatics, it is calculated as follows -

\begin{itemize}
    \item Calculate the $\log_2$ of the gene expression value, \textbf{cdRPKM}, for each timepoint. Let this be $x_i$.
    \item Subtract $x_0$ (the value at timepoint 0) from each successive timepoint $x_i$. 
    \item Add this as a new attribute to the RNASeq and RPF data.
\end{itemize}

For TE data, since the values are already log ratios, we merely subtract each timepoint's value from the first timepoint.

Even with the new features, we might encounter problems due to \textit{noise}. Fortunately, we have a biological approach of dealing with this. 

When a cell is infected with a virus, the RNASeq does not change much (as it measures the total content of all genes) while those genes that help in virus production will get translated more, i.e. the RPF of these genes will increase. Since TE measures translational efficiency ($TE = \frac{RPF}{RNASeq}$), the TE of such genes will also keep increasing across timepoints. 

We filter out those genes that do not change much at the RNASeq level. We retain genes which have \textbf{foldChange} values for each timepoint between $-0.5$ and $0.5$. More formally, if $x_i$ is the foldChange for $1 \leq i \leq 5$, then $-0.5 \leq x_i \leq 0.5 $ for each $x_i$ (later on, we reduce this threshold further to [-0.3,0.3] to check whether the GO terms we obtained are indeed important or not).

After this filtering, we are left with \textbf{5576 genes} that are highly translated.

As we are interested in genes that are highly translated between timepoints, we apply hierarchical clustering only on the \textbf{foldChange} features in the TE dataset. We plot heatmaps for different linkages, and check the general trends to find the number of clusters. In a heatmap, the values of each gene are plotted according to the colour key (bottom left). 

The values in the heatmaps are sorted accorded to how they have been clustered.

\begin{figure}[h]
    \centering
    \includegraphics[width=0.4\textwidth]{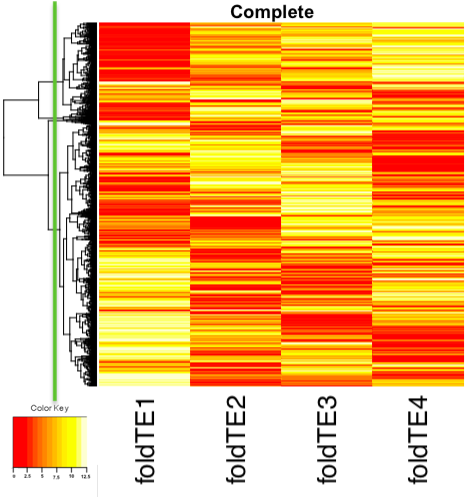}
    \caption{Heatmap with Complete Linkage Hierarchical Clustering for Filtered Data}
    \label{fig:IdealClustering}
\end{figure}

In our case, we find that \textit{complete} linkage performs the best. We can identify 4-6 distinct clusters from the heatmap, and cut the dendogram accordingly (represented by the green line). The result is shown in Figure \ref{fig:IdealClustering}.

After applying hierarchical clustering and doing heatmap analysis on the filtered data, we infer that $K$ lies between 4 and 6. We use these values of $K$ for clustering for Step 5 of Procedure \ref{Analysis}.

\subsection{Model-Based Clustering Results}

\subsubsection{K-Means}
We first apply K-Means on all the datasets to determine the optimal $K$.

\begin{figure}[h]
    \centering
    \begin{subfigure}[b]{0.35\textwidth}
        \includegraphics[width=\textwidth]{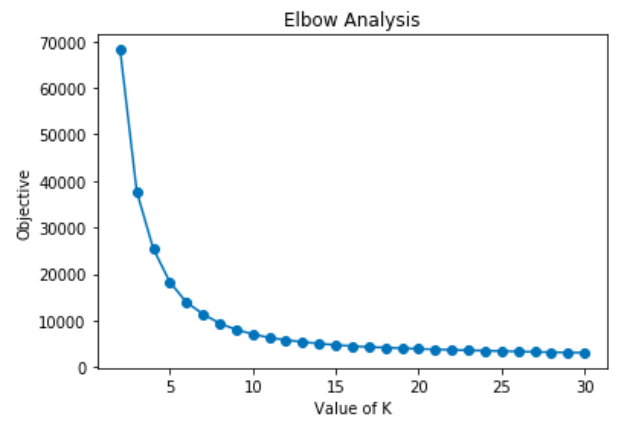}
        \caption{Elbow Method}
    \end{subfigure}
    \begin{subfigure}[b]{0.35\textwidth}
        \includegraphics[width=\textwidth]{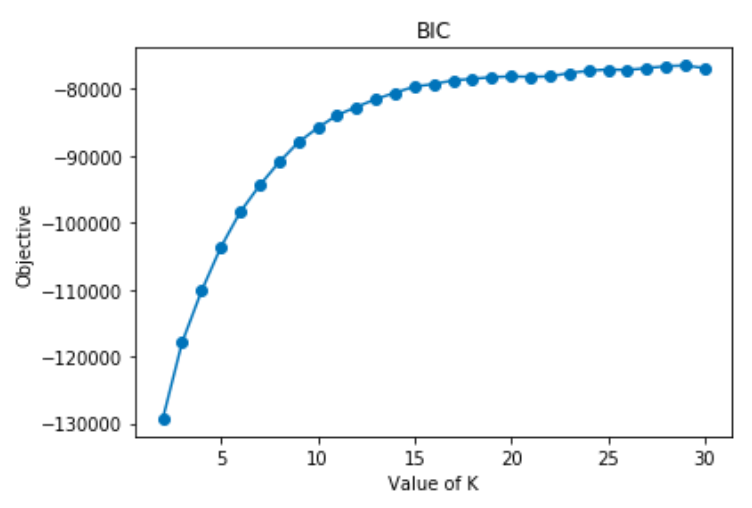}
        \caption{BIC Curve}
    \end{subfigure}
    \caption{Cluster Analysis for RNASeq}
    \label{fig:KMeanseval}
\end{figure}

We initialise the algorithm using K-Means++ for optimal clusterings, and apply the Elbow method and BIC to evaluate our results. Results for RNASeq are given in Figure \ref{fig:KMeanseval}.

The full set of graphs for all datasets are given in Appendix \ref{app:A}, Section \ref{Kmeansfigures}.

The results are summarised in Table \ref{tab:KMeanstable}. "-" means inconclusive.
\begin{table}[h]
    \centering
    \begin{tabular}{| m{5cm} | m{3cm} | m{3cm} |}
    \hline
         \textbf{Data} & \textbf{Elbow Method} & \textbf{BIC}   \\\hline
         RNASeq (5 points) & - &  15*\\\hline 
         RPF (5 points) & - & - \\\hline
         TE (5 points) & 5 & 17* \\\hline
    \end{tabular}
    \caption{Optimal K for different datasets, K-Means}
    \label{tab:KMeanstable}
\end{table}

{\tiny *For BIC, if there is no clear peak, we take the first local maxima as $K$}

From the plots, we observe that determining the optimal $K$ is difficult.

There is no clear 'elbow' point, and the BIC keeps increasing with $K$; thus K-Means is not penalising enough. (Using AIC is futile, as the penalisation would be even lower in this case). The elbow plots and GO analysis do not mutually agree on a $K$. 

Despite the mixed results from the information criteria metrics, we went ahead and performed GO analysis as per the $K$ determined in Table \ref{tab:KMeanstable}. However, the GO results were also found to be inconclusive - there were multiple clusters associated with the \textit{'viral process'} term, and the p-value was quite high (i.e. the terms were not enriched).

Thus, there is no reliable estimate for $K$. This is most likely due to 2 reasons -
\begin{itemize}
    \item K-means cannot capture temporal dependencies in data.
    \item Data is too noisy.
\end{itemize}

Both can be solved through feature engineering - we add the changes between the timepoints as features. 

We filter the data as described in Section \ref{sec:ExpAnal}, and using the heatmap results, perform K-Means clustering for 4,5 and 6 clusters on the \textbf{foldChange} TE data.

We plot heatmaps to observe the general trends of clusters from this heuristic method, and identify those clusters which have an increasing trend (the clusters are plotted top to bottom in the heatmap and demarcated by black lines).

\begin{figure}[h]
    \centering
    \begin{subfigure}[b]{0.3\textwidth}
        \includegraphics[width=\textwidth]{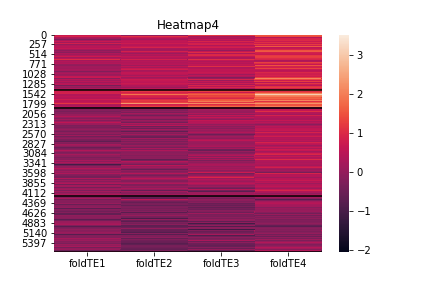}
        \caption{4 clusters}
        \label{a}
    \end{subfigure}
    \begin{subfigure}[b]{0.3\textwidth}
        \includegraphics[width=\textwidth]{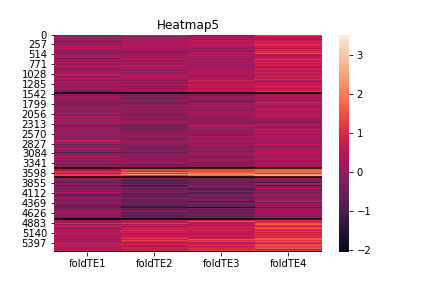}
        \caption{5 clusters}
        \label{b}
    \end{subfigure}
    \begin{subfigure}[b]{0.3\textwidth}
        \includegraphics[width=\textwidth]{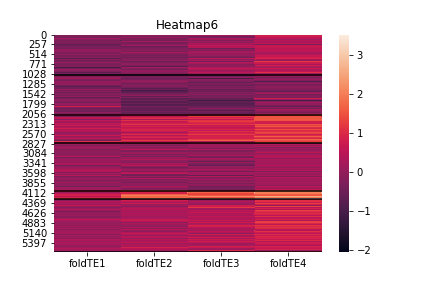}
        \caption{6 clusters}
        \label{c}
    \end{subfigure}
    \caption{Heatmaps for K-Means clustering on TE data (top to bottom) }
\end{figure}

Cluster 2, 3 and 5 in Figures \ref{a}, \ref{b} and \ref{c} have an increasing trend, so we focus on the GO analyses of those clusters (covered in Section \ref{Bio}).

This heuristic method gives us GO results (specifically, terms related to GTPases from Section \ref{GTPAse}) that are relevant to our hypothesis. This is covered in greater detail in Section \ref{Bio}.

\subsubsection{Gaussian Mixture Models}

As seen in the K-Means approach, it is futile to do GO analysis on complete datasets (especially for the noisy RNASeq data, as we will most likely observe no significant trends by merely tracking overall changes in gene expression).

To test this, we still apply GMMs on the complete TE data (which are most likely to show certain trends due to genes being translated). As expected, we have no definitive results. The GO is unclear, and the BIC keeps increasing with $K$ for all covariance settings, offering no clear peak. The plots are in Appendix \ref{app:A}, Section \ref{GMMFigures}.

We decide to apply GMMs on the feature-engineered data from Section \ref{FeatureEngineering}. We use the full covariance matrix and do BIC analysis for $2 \leq K \leq 20$. 

\begin{figure}[h]
    \centering
    \includegraphics[width=0.5\textwidth]{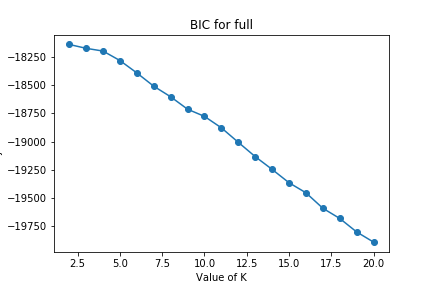}
    \caption{BIC for full covariance matrix, TE data}
    \label{BICfinal}
\end{figure}

For filtered TE data, we conclude that the ideal number of clusters are $K=2$ from Figure \ref{BICfinal}. We plot the heatmap and general foldChange graph (in Section \ref{TrendPlotsGMM}) for this clustering to check the general trend as per the heatmap (clusters are demarcated by the green line).

\begin{figure}[h]
    \centering
    \includegraphics[width=0.5\textwidth]{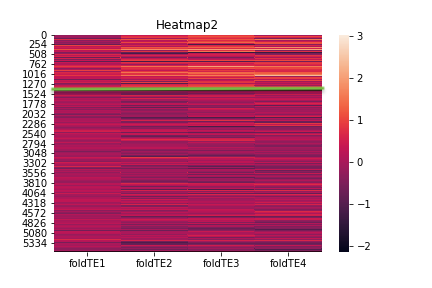}
    \caption{Heatmap for 2 clusters, GMM clustering on TE data}
    \label{GMMHeatmap}
\end{figure}

We see in Figure \ref{GMMHeatmap} that Cluster 1 has an increasing trend. The GO analysis of this cluster is similar to that obtained through K-Means - we again observe terms related to GTPases.

We still perform GMM-based clustering for $K=4,5,6$ for evaluating GMMs performance against the benchmark algorithm in Section \ref{Performance}.

\subsubsection{HMM-based Clustering }

We speculated that HMM-based clustering might perform better than other algorithms on the complete and unfiltered datasets, as HMMs inherently take the time dependencies into account. 

We assume $3$ hidden states for simplicity, and apply HMM-based clustering on all the datasets. The plots are in Section \ref{HMMBIC}, while the results are summarised in Table \ref{tab:HMMTable}.

\begin{table}[h]
    \centering
    \begin{tabular}{| m{5cm}| m{5cm} |}
    \hline
         \textbf{Data} &  \textbf{BIC}   \\\hline
         RNASeq (5 points) &  13*\\\hline 
         RPF (5 points) & 8*  \\\hline
         TE (5 points) &  16* \\\hline
    \end{tabular}
    \caption{Optimal K for different datasets, HMMs}
    \label{tab:HMMTable}
\end{table}

{\tiny *For BIC, if there is no clear peak, we take the first local maxima as $K$}

Again, we observe no clear peaks. The GO results for the local peaks are inconclusive, although some of the terminologies in the GO for RPF and TE do include GTPases and other related terms. However, there were no other clear trends.

We can deduce from BIC's inability to determine an optimal $K$ that we may be estimating too many parameters and unnecessarily complicating the models (Occam's Razor).

\begin{figure}[h]
    \centering
    \begin{subfigure}[b]{0.4\textwidth}
        \includegraphics[width=\textwidth]{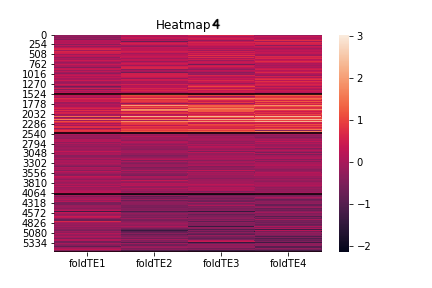}
        \caption{Heatmap for 4 clusters}
    \end{subfigure}
    \begin{subfigure}[b]{0.4\textwidth}
        \includegraphics[width=\textwidth]{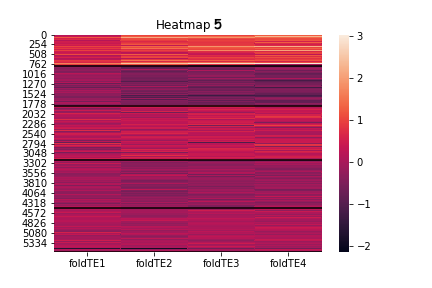}
        \caption{Heatmap for 5 clusters}
    \end{subfigure}
    \caption{HMM Heatmap results on feature-engineered data}
    \label{fig:HMMHeatmap}
\end{figure}

We proceed to the next steps of the analysis. 
We apply HMM-based clustering on the feature-engineered data for $K=4,5$ (we skip 6 as HMMs take the longest time to train) and generate heatmaps in Figure \ref{fig:HMMHeatmap}. We then check the heatmaps for up-regulated clusters, and perform GO analysis on the Clusters 2 and 1 respectively from Figure \ref{fig:HMMHeatmap}.

The trend graphs of each cluster are given in Section \ref{HMMTrends}.

The GO results from HMMs are comparable to those from GMMs and K-Means, and we again observe GTPases in the gene ontology of the up-regulated clusters.

\subsubsection{Bayesian Networks}

A limitation of Bayesian networks is that the current implementation is only applicable to discrete random variables (or discrete features).

Although there is some literature that deals with hybrid Bayesian Networks that model data containing both discrete and continuous attributes \cite{cobb2007bayesian}, data with purely continuous variables are difficult to model using Bayesian Networks.

Thus, as a first step, we \textit{discretise} our gene expression data \cite{pham2009unsupervised}.

According to \cite{pham2009unsupervised}, there is no conclusive way of determining $K$ for CL Multinets based clustering, so we skip AIC/BIC analysis here. Instead, we perform clustering using $K=4,5,6$ to determine the structure of the underlying Bayesian network for the complete datasets. 

The heatmap results for the TE dataset showed some increasing trends (covered in Section \ref{fig:BayesianNetworksHeatmap}; the GO analysis for the complete TE dataset included GTPases in more abundance than HMM-based clustering.

\begin{figure}[h]
  \begin{center}
    \includegraphics[width=0.15\textwidth]{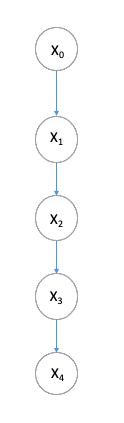}
  \end{center}
    \caption{Tree Structure}
  \label{fig:tree}
\end{figure}

Moreover, the underlying structure for each tree (determined by the clustering, for different $K$) unfailingly resulted in the same line graph (in Figure \ref{fig:tree}), with the timepoints in order (timepoint 1, followed by timepoint 2, and so on). This, along with the GO results, shows that Bayesian networks cluster even noisy data with a certain level of accuracy. 

\setlength\intextsep{0pt}
\setlength\intextsep{0pt}

We then apply Bayesian networks to the feature engineered TE dataset. The expression values range from -2 to 4, and we choose 100 bins between -2 and 4, given by $[-2.0, -1.94,...,3.94,4.0]$ and then digitise each observation according to the bins it belongs to.
 
\begin{figure}[h]
    \centering
    \begin{subfigure}[b]{0.4\textwidth}
        \includegraphics[width=\textwidth]{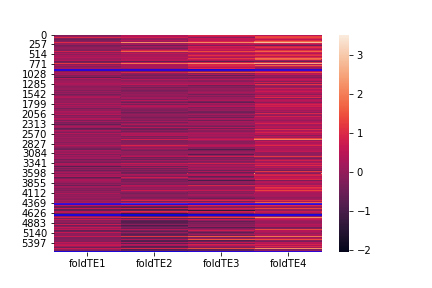}
        \caption{Heatmap for 4 clusters}
    \end{subfigure}
    \begin{subfigure}[b]{0.4\textwidth}
        \includegraphics[width=\textwidth]{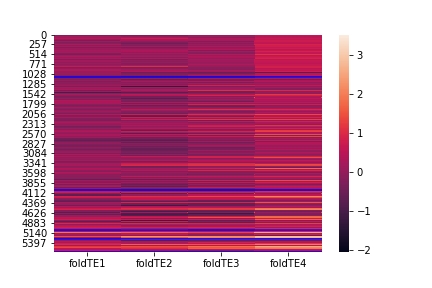}
        \caption{Heatmap for 5 clusters}
    \end{subfigure}
        \begin{subfigure}[b]{0.4\textwidth}
        \includegraphics[width=\textwidth]{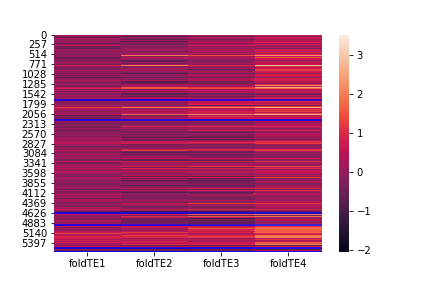}
        \caption{Heatmap for 6 clusters}
    \end{subfigure}
    \caption{CL Multinets Heatmap results on feature-engineered data}
    \label{fig:BNHeatmap}
\end{figure}

The heatmaps for 6 clusters are given in Figure \ref{fig:BNHeatmap}. The blue lines demarcate the clusters.

Although the heatmaps are not as uniform and do not display concrete trends (as those from K-Means clustering) the GO results are reliable and contain more GTPase-related terms, along with some new terms related to viral translation (e.g. G2/M transition of mitotic cell \cite{davy2007g2}).

\subsection{Performance Evaluation}
\label{Performance}

We compare algorithms against a benchmark for the feature engineered TE dataset (with the foldChange features).

Bayesian networks consider the inherent time-dependency between attributes, resulting in a graph with the timepoints in correct sequence each time. 

The GO results are also significant as they capture those genes that are up-regulated yet also distinct from each other despite having similar expression values (unlike K-Means, which clusters according to Euclidean distance between absolute attribute values). 

We pick Bayesian networks as our benchmark algorithm, and treat the results obtained from it as the \textit{ground truth}. 

We use the ME distance metric to compare the other techniques against Bayesian networks for $K=4,5,6$. For ME Distance, the closer it is to 0, the better the clustering is. We calculate the accuracy as 100(1-$d_{ME}$)\%.

\begin{table}
    \centering
    \begin{tabular}{|p{3cm}|p{3cm}|p{3cm}|p{3cm}|}
    \hline
           & \multicolumn{3}{|c|}{\textbf{Accuracy for Clusters}}  \\\hline
         \textbf{Algorithm} &  \hspace{1.35cm}\textbf{4} & \hspace{1.35cm}\textbf{5} & \hspace{1.35cm}\textbf{6} \\\hline 
         K-Means & \hspace{0.95cm}56.52\%  & \hspace{0.95cm}48.35\% & \hspace{0.95cm}41.61\% \\\hline
         GMMs & \hspace{0.95cm}58.40\% & \hspace{0.95cm}53.11\%  & \hspace{0.95cm}50.31\%\\\hline
         HMMs & \hspace{0.95cm}68.72\%  & \hspace{0.95cm}79.45\% & \hspace{1.45cm}-\\\hline
    \end{tabular}
    \caption{Performance evaluation for all algorithms*}
    \label{tab:Comparison4}
\end{table}
{\tiny *compared to Bayesian networks}

\subsection{Biological Results}
\label{Bio}

Existing literature on picornavirus (which the EV-A71 virus is a type of) suggests that the proteins in the virus interact with certain GTPases and disrupt normal cellular functions (like nuclear transport of other viral and nuclear proteins).
\cite{porter2006picornavirus}. 

Looking more closely at the collection of heatmaps obtained by all the algorithms, we identify all those clusters that are up-regulated and contain terms containing/related to GTPases in their GO analysis.
\begin{figure}[h]
    \includegraphics[width=1.0\textwidth]{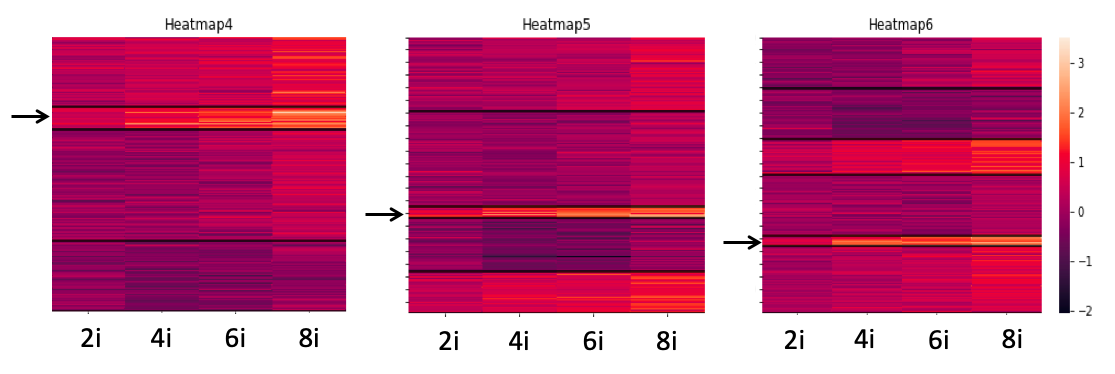}
    \caption{Heatmap from K-Means, RNASeq filter [-0.5, 0.5]}
    \label{fig:final1}
\end{figure}

To be more rigorous about the terms we observe, we also reduce the filter to $[-0.3, 0.3]$ and check the GO analysis of the up-regulated clusters. For each cluster, we only consider enriched GO terms (those with low p-values, mentioned in Section \ref{PValue}).

\begin{figure}[h]
    \includegraphics[width=1.0\textwidth]{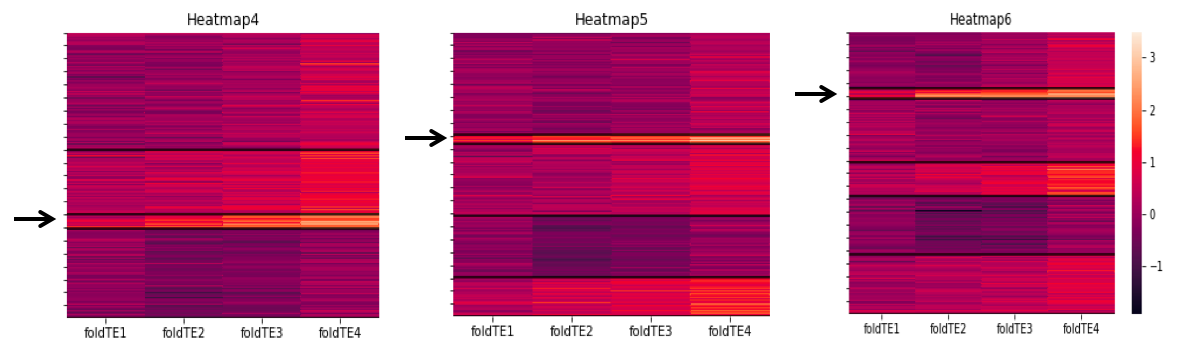}
    \caption{Heatmap from K-Means, RNASeq filter [-0.3, 0.3]}
    \label{fig:final2}
\end{figure}

Throughout our (repeated) experiments, we find that terms related to GTPases from \textit{Biological Processes} (BP) and \textit{Molecular Functions} (MF) keep re-appearing. The terms of interest are -
\begin{itemize}
    \item Positive regulation of GTPase activity
    \item Activation of GTPase activity
    \item Regulation of small GTPase mediated signal transduction
    \item Positive regulation of transcription, DNA templated
    \item Signal transduction
\end{itemize}

Through the principle of guilt-by-association mentioned in Section \ref{sec:Results}, we also identify other enriched terms that are of interest -
\begin{itemize}
    \item Viral Process
    \item DNA repair
    \item Negative regulation of transcription from RNA pol II promoter
    \item Establishment of cell polarity
    \item Smoothened signaling pathway
    \item Positive regulation of dendrite development
    \item G2/M transition of mitotic cell
\end{itemize}

Some GO terms (viral process, regulation of small GTPase mediated signal transduction, DNA repair) are observed across multiple clusters in the HMMs and Bayesian Networks analysis. This suggests interactions between the clusters too.

From the GO analysis, we find plenty of evidence from our datasets to support our hypothesis that a certain class/member of GTPase is important for establishing this infection.

% SEC 7
\clearpage
\section{Conclusion}

This thesis explored the application of unsupervised learning techniques on viral genomic data - more specifically, human cells infected with the EV-A71 virus (which causes HFMD). We attempted to determine host factors important to the propagation of this virus, and identified several key biological terms based on our clustering results.

\subsection{Summary}
As described in Section \ref{experimentbio}, we collected RNASeq and RPF gene expression data at different timepoints.
In Figure \ref{fig:virus}, we saw that the space occupied by the virus increases with time. Initial analysis of the complete datasets (RNASeq, RPF, TE) were consistent with this trend. Next, we attempted to determine groups of genes crucial to viral translation. Initial clustering results contained some relevant GO terms (e.g. viral process, DNA repair) that appeared in multiple clusters; however, the trend graphs and heatmaps proved to be inconclusive, suggesting that our datasets were too \textit{noisy}, containing many points insignificant to our analysis.

We then focused on those genes which were highly translated at the RPF level but not at the RNASeq level. Collating data from all our GO analyses, we identified the following terms - Positive regulation of GTPase activity, Activation of GTPase activity, Regulation of small GTPase mediated signal transduction, Positive regulation of transcription, DNA templated and Signal transduction.

These terms imply that some proteins in the class GTPase are definitely important to viral translation, possibly increasing the production of viral genes at key phases.

Following up on these results, we conduct more extensive literature reviews to verify the link between small GTpases and picornavirus infection. 

\subsubsection{Biological Experiments}
We also design experiments to verify whether GTPases are relevant to the EV-A71 virus.

The first experiment tracks changes in GTPase activity throughout infection, focusing more on Rho GTPases. For GTPase protein-specific activity, there are also targeted studies to check for specificity of GAP/GEF, which are related to activation of GTPases.

The second experiment inhibits specific GTPases to check their effect on viral cells.
3 kinds of drugs targeting different GTPases are used - 
\begin{itemize}
    \item Specific Rho drugs e.g. Rhosin or C3 transferase against RhoA/B/C, but not cdc42/Rac1
    \item Specific cdc42 drugs e.g. Casin, ML141
    \item Specific Rho kinase inhibitors e.g. Rhodblock series
\end{itemize}

{\large Thus, the appearance of terms related to viral process in the GO analysis for all methods suggests that our clustering was useful. The abundance of GTPase-related terms also support our hypothesis that they are indeed vital to the propagation of the virus inside the cell.}

\section{Future Work}
\label{sec:FutWork}

\subsection{Algorithms}

\subsubsection{Tuning}
We can improve the performance of our current model-based clustering techniques by incorporating additional data or adjusting the hyperparameters for the prior distributions we use for initialisation. We can consider discretising the data for Hidden Markov Model based clustering \cite{zhao2010hmm}; conversely, we can consider continuous variables for Bayesian networks \cite{hofmann1996discovering}.

\subsubsection{Different Algorithms}
We can consider self-organizing maps (SOMs) for clustering gene expression data too \cite{tamayo1999interpreting}. SOMs impose partial structure on the clusters (in contrast to the rigid structure of hierarchical clustering, the strong prior hypotheses used in Bayesian clustering, and the non-structure of k-means clustering) and facilitate easy visualization and interpretation. SOMs have good computational properties and are easy to implement, reasonably fast, and scalable to large data sets too.

We can also consider soft clustering, or fuzzy clustering; where one gene may belong to more than one cluster. Existing fuzzy-clustering techniques can indeed be modified for short time-series gene expression data \cite{moller2003fuzzy}.

%%%%%%%%%%%%%%%%%%%%%%%%%%%%%%%%%%%%%%%%%%%%%%%%%%%%%%%%%%%%%
%APPENDICES
%%%%%%%%%%%%%%%%%%%%%%%%%%%%%%%%%%%%%%%%%%%%%%%%%%%%%%%%%%%%%

\appendix
\renewcommand*{\thesection}{\Alph{section}}\textbf{}

% APPENDIX A
\clearpage
\section{Appendix}
\label{app:A}
\subsection{K-Means} 
\subsubsection{BIC Plots}
\label{Kmeansfigures}
\begin{figure}[h]
    \centering
    \begin{subfigure}[b]{0.3\textwidth}
        \includegraphics[width=\textwidth]{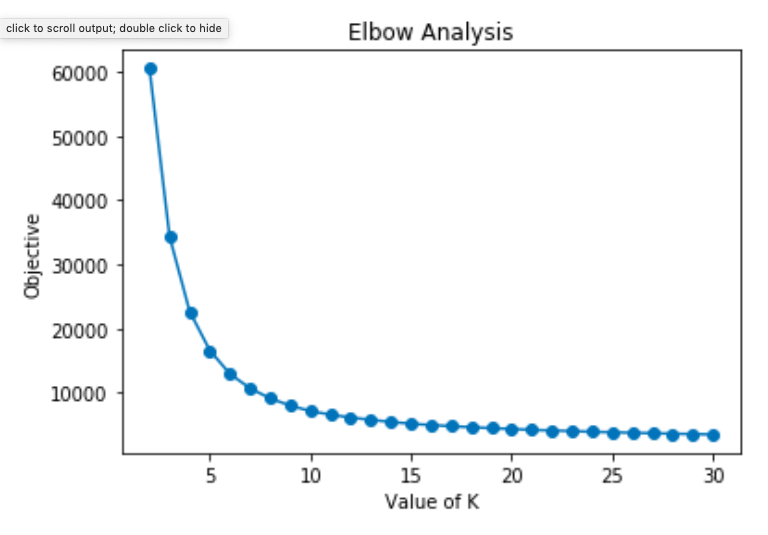}
        \caption{Elbow Method}
    \end{subfigure}
    \begin{subfigure}[b]{0.3\textwidth}
        \includegraphics[width=\textwidth]{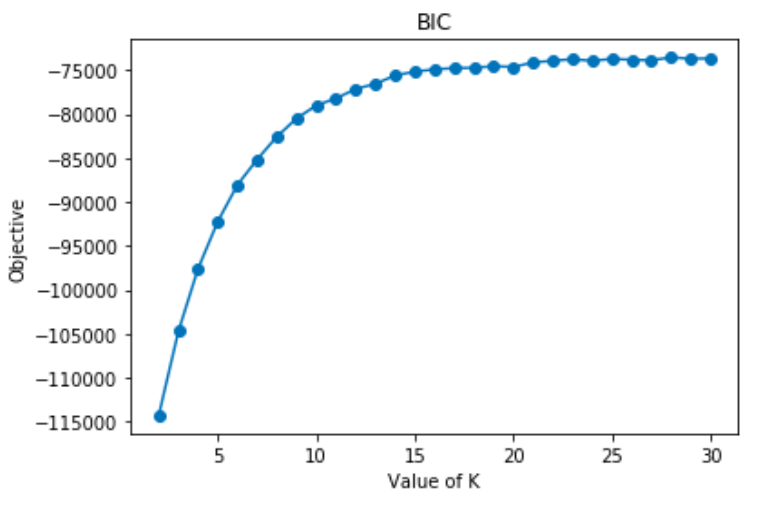}
        \caption{BIC Curve}
    \end{subfigure}
    \caption{Cluster Analysis for RPF}
\end{figure}

\begin{figure}[h]
    \centering
    \begin{subfigure}[b]{0.3\textwidth}
        \includegraphics[width=\textwidth]{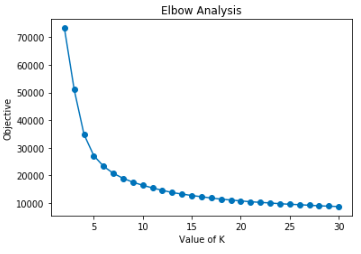}
        \caption{Elbow Method}
    \end{subfigure}
    \begin{subfigure}[b]{0.3\textwidth}
        \includegraphics[width=\textwidth]{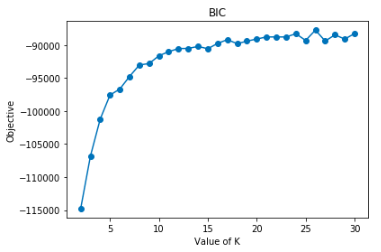}
        \caption{BIC Curve}
    \end{subfigure}
    \caption{Cluster Analysis for TE}
\end{figure}

\subsection{Gaussian Mixture Models}
\label{GMMFigures}

\subsubsection{BIC Plots}
\begin{figure}[h]
    \centering
    \begin{subfigure}[b]{0.3\textwidth}
        \includegraphics[width=\textwidth]{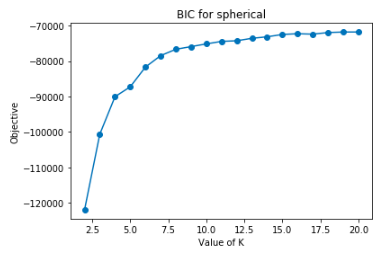}
        \caption{Spherical}
    \end{subfigure}
    \begin{subfigure}[b]{0.3\textwidth}
        \includegraphics[width=\textwidth]{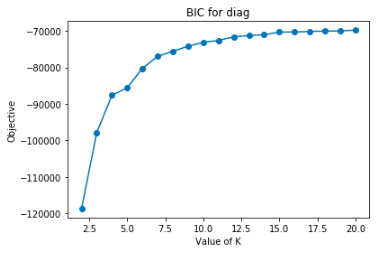}
        \caption{Diagonal}
    \end{subfigure}
    \begin{subfigure}[b]{0.3\textwidth}
        \includegraphics[width=\textwidth]{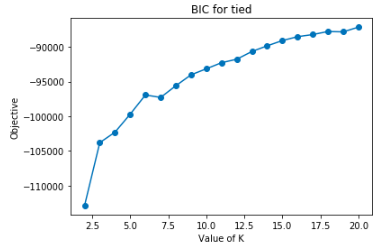}
        \caption{Tied}
    \end{subfigure}
    \begin{subfigure}[b]{0.3\textwidth}
        \includegraphics[width=\textwidth]{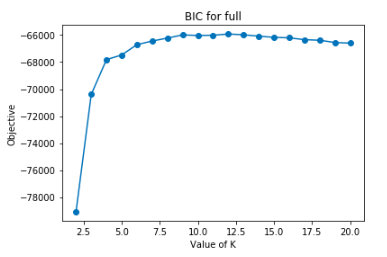}
        \caption{Full}
    \end{subfigure}
    \caption{BIC for different covariance settings in GMMs, TE}
\end{figure}

\subsubsection{Trend Plots}
\label{TrendPlotsGMM}
\begin{figure}[h]
    \centering
    \begin{subfigure}[b]{0.4\textwidth}
        \includegraphics[width=\textwidth]{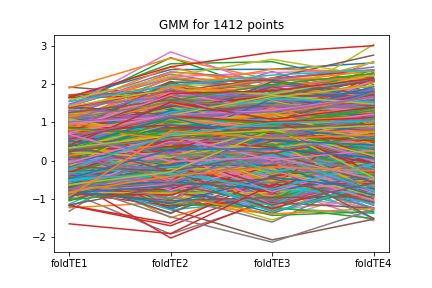}
        \caption{Cluster 1}
    \end{subfigure}
    \begin{subfigure}[b]{0.4\textwidth}
        \includegraphics[width=\textwidth]{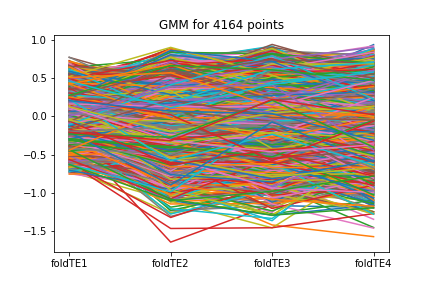}
        \caption{Cluster 2}
    \end{subfigure}
    \caption{Trend plots for foldChange values, 2 clusters, GMM}
\end{figure}

\subsection{Bayesian Networks}
\subsubsection{Heatmaps for Complete Datasets}

The clusters are demarcated by red lines.
\begin{figure}[h]
    \centering
    \begin{subfigure}[b]{0.45\textwidth}
        \includegraphics[width=\textwidth]{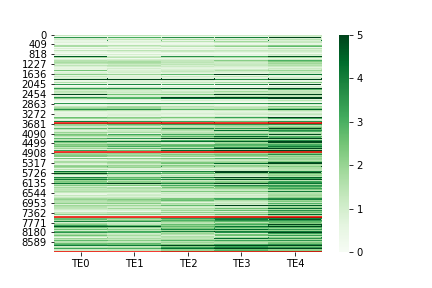}
        \caption{4 clusters}
    \end{subfigure}
    \begin{subfigure}[b]{0.45\textwidth}
        \includegraphics[width=\textwidth]{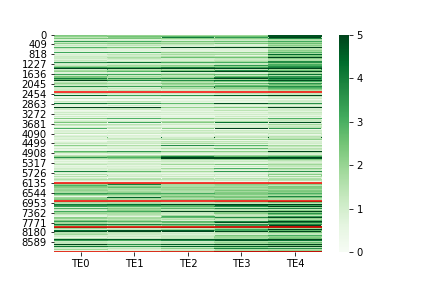}
        \caption{5 clusters}
    \end{subfigure}
    \begin{subfigure}[b]{0.45\textwidth}
        \includegraphics[width=\textwidth]{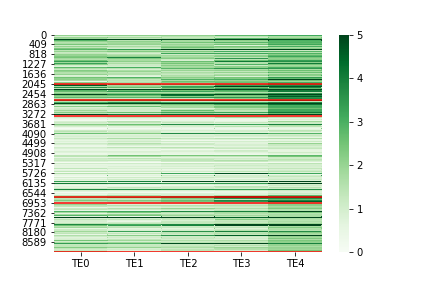}
        \caption{6 clusters}
    \end{subfigure}
    \caption{Heatmaps for foldChange TE value}
    \label{fig:BayesianNetworksHeatmap}
\end{figure}

\subsection{HMM-based Clustering}
\subsubsection{BIC Plots}
\label{HMMBIC}

\begin{figure}[h]
    \centering
    \begin{subfigure}[b]{0.3\textwidth}
        \includegraphics[width=\textwidth]{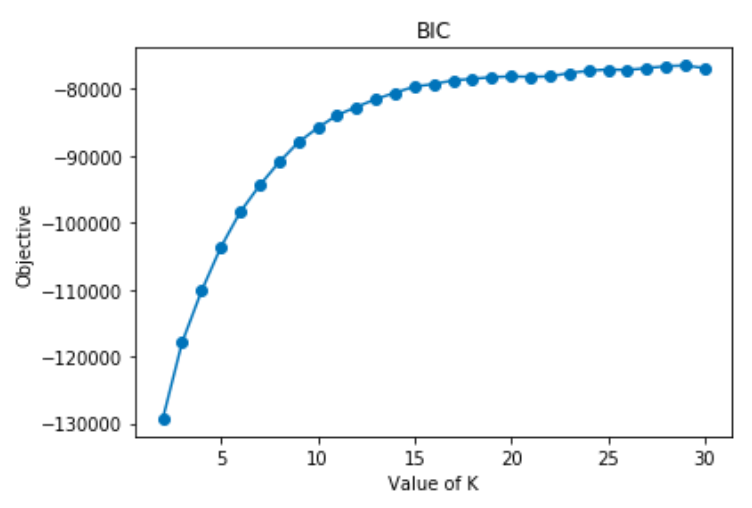}
        \caption{RNASeq data}
    \end{subfigure}
    \begin{subfigure}[b]{0.3\textwidth}
        \includegraphics[width=\textwidth]{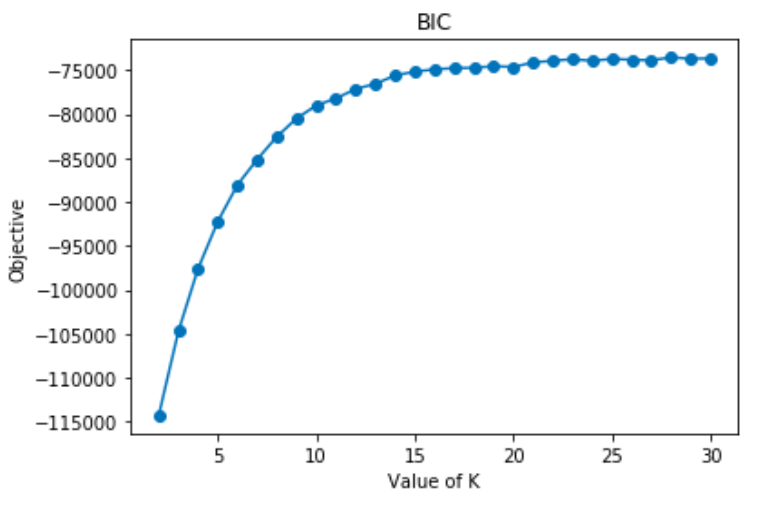}
        \caption{RPF Data}
    \end{subfigure}
    \begin{subfigure}[b]{0.3\textwidth}
        \includegraphics[width=\textwidth]{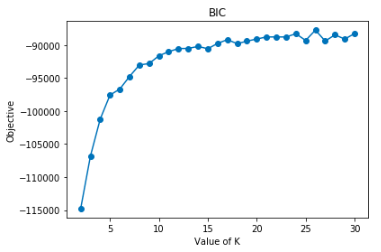}
        \caption{TE Data}
    \end{subfigure}
    \caption{BIC results for HMM-based clustering}
\end{figure}

\subsubsection{Trend plots}
\label{HMMTrends}
\begin{figure}[h]
    \centering
    \begin{subfigure}[b]{0.35\textwidth}
        \includegraphics[width=\textwidth]{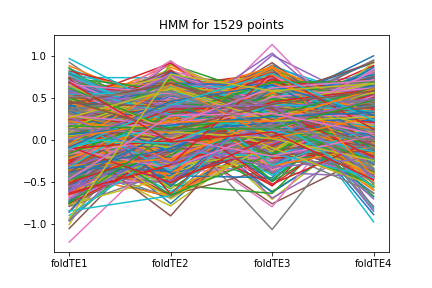}
        \caption{Cluster 1}
    \end{subfigure}
    \begin{subfigure}[b]{0.35\textwidth}
        \includegraphics[width=\textwidth]{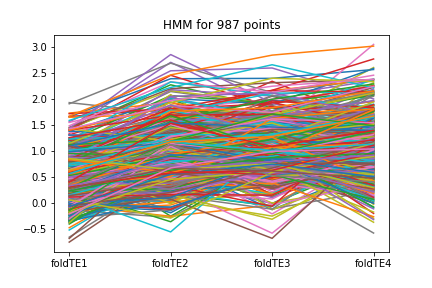}
        \caption{Cluster 2}
    \end{subfigure}
    \begin{subfigure}[b]{0.35\textwidth}
        \includegraphics[width=\textwidth]{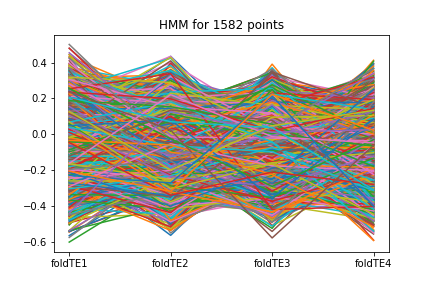}
        \caption{Cluster 3}
    \end{subfigure}
    \begin{subfigure}[b]{0.35\textwidth}
        \includegraphics[width=\textwidth]{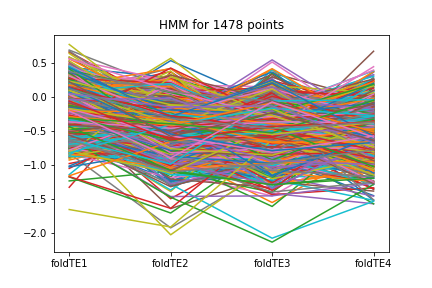}
        \caption{Cluster 4}
    \end{subfigure}
    \caption{Trend plots for foldChange values, 4 clusters, HMM-based clustering}
\end{figure}

\begin{figure}[h]
    \centering
    \begin{subfigure}[b]{0.35\textwidth}
        \includegraphics[width=\textwidth]{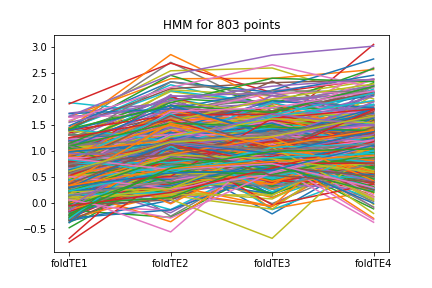}
        \caption{Cluster 1}
    \end{subfigure}
    \begin{subfigure}[b]{0.35\textwidth}
        \includegraphics[width=\textwidth]{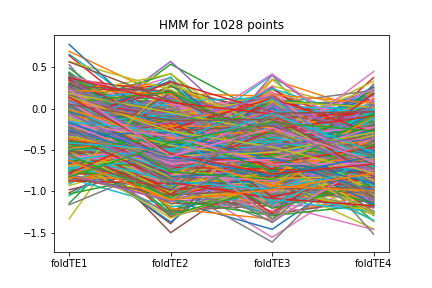}
        \caption{Cluster 2}
    \end{subfigure}
    \begin{subfigure}[b]{0.35\textwidth}
        \includegraphics[width=\textwidth]{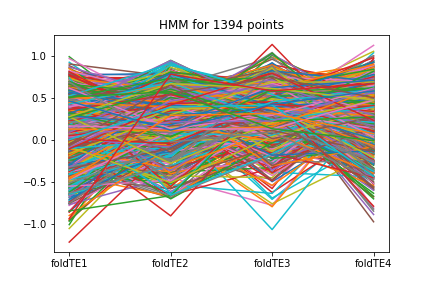}
        \caption{Cluster 3}
    \end{subfigure}
    \begin{subfigure}[b]{0.35\textwidth}
        \includegraphics[width=\textwidth]{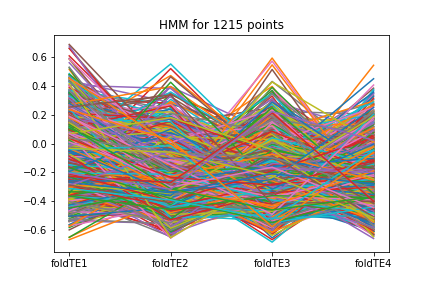}
        \caption{Cluster 4}
    \end{subfigure}
        \begin{subfigure}[b]{0.35\textwidth}
        \includegraphics[width=\textwidth]{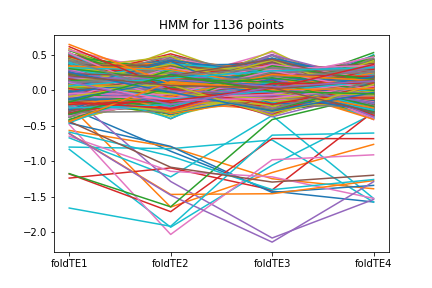}
        \caption{Cluster 5}
    \end{subfigure}
    \caption{Trend plots for foldChange values, 5 clusters, HMM-based clustering}
\end{figure}

%%%%%%%%%%%%%%%%%%%%%%%%%%%%%%%%%%%%%%%%%%%%%%%%%%%%%%%%%%%%%
%BIBLIOGRAPHY
%%%%%%%%%%%%%%%%%%%%%%%%%%%%%%%%%%%%%%%%%%%%%%%%%%%%%%%%%%%%%

\clearpage
\renewcommand*{\thesection}{}\textbf{}

\listoffigures
\clearpage
\listoftables
\clearpage

\bibliographystyle{plain}
\bibliography{Bibliography.bib}

\end{document}